\documentclass[onecolumn,draftclsnofoot, 12pt, ]{IEEEtran}

\usepackage{times}
\usepackage[T1]{fontenc}
\usepackage{algorithm}
\usepackage{algorithmicx}
\usepackage{algpseudocode}
\usepackage{amsthm}
\usepackage{graphicx}
\usepackage{amsmath}
\usepackage{setspace}
\usepackage{amsfonts}
\usepackage{amsbsy}
\usepackage{cite}      

\usepackage{graphicx}  

\usepackage{subfigure} 

\usepackage{amsmath}   
\usepackage{array}
\newtheorem{theorem}{Theorem}

\newtheorem{corollary}[theorem]{Corollary}

\begin{document}
%
\title{Millimeter Wave Massive MIMO Downlink Per-Group Communications with Hybrid Linear Precoding }

\author{T.~Ketseoglou,~\IEEEmembership{Senior Member,~IEEE}, M.C.~Valenti,~\IEEEmembership{Fellow,~IEEE}, and E.~Ayanoglu,~\IEEEmembership{Fellow,~IEEE}  %
\thanks{T. Ketseoglou is with the Electrical and Computer Engineering Department, California State Polytechnic University, Pomona, USA (e-mail: tketseoglou@cpp.edu). M. Valenti is with the Lane Department of Computer Science and Electrical Engineering, West Virginia University, Morgantown, USA (email:  mvalenti@wvu.edu). E. Ayanoglu is with the Electrical Engineering and Computer Science Department, University of California, Irvine, USA (e-mail:ayanoglu@uci.edu). This work was partially supported by NSF grant 1547155.}}


\maketitle

\begin{abstract}
We address the problem of analyzing and classifying in groups the downlink channel environment in a millimeter-wavelength cell, accounting for path loss, multipath fading, and User Equipment (UE) blocking, by employing a hybrid propagation and multipath fading model, thus using accurate inter-group interference modeling. The base station (BS) employs a large Uniform Planar Array (UPA) to facilitate massive Multiple-Input, Multiple-Output (MIMO) communications with high efficiency. UEs are equipped with a single antenna and are distributed uniformly within the cell. The key problem is analyzing and defining groups toward precoding. Because balanced throughput is desired between groups, Combined Frequency and Spatial Division and Multiplexing (CFSDM) is found to be necessary. We show that by employing three or four subcarrier frequencies, depending on the number of UEs in the cell and the cell range, the UEs can be efficiently separated into high throughput groups, with each group employing Virtual Channel Model Beams (VCMB) based inner precoding, followed by efficient Multi-User Multiple-Input Multiple-Output (MU-MIMO) outer precoders. For each group, we study three different sub-grouping methods offering different advantages. We show that the improvement offered by Zero-Forcing Per-Group Precoding (ZF-PGP) over Zero-Forcing Precoding (ZFP) is  very high. In addition, for medium-correlation channels, it is shown that ZF-PGP performance is near to the ideal one offered by a Virtual Additional Antenna Concept PGP (VAAC-PGP). Finally, a new technique for power allocation among different Per-Group Precoding within Groups (PGP-WG) groups is proposed, called Optimized Per-Group Power Allocation (OPGPA), which allows for high power efficiency with equal throughput among all UEs.

\end{abstract}

\section{{Introduction}}
Because it offers a wide spectrum that can support short-range high-rate wireless connectivity \cite{Rapp_sr}, millimeter-wavelength communication is an attractive solution for future wireless applications, including massive Multiple-Input Multiple-Output (MIMO) applications\cite{Marzetta1,Marzetta2,Marzetta3, Swindlehurst}.
Downlink input-output mutual information maximizing (IOMIM) linear precoding with finite-alphabet inputs, e.g., by employing Quadrature Amplitude Modulation (QAM), has been extensively studied \cite{TK_EA_QAM,Schoeber_THP_TWC,Xiao, TK_EA_TCOM2, LozanoA, Yongpeng, Xiao2, SCSI} due to its potential to offer high data rate collectively. However, all existing studies have focused on multipath fading without considering essential propagation effects in mmWave communications such as User Equipment (UE) blocking, path loss, and varying fading scaling factors. The latter effects have been modeled in \cite{MV1,MV2} for mmWave device-to-device communications  in a flexible way with success.

In this paper, we apply the model proposed in \cite{MV1,MV2} for capturing mmWave propagation effects and combine it with UE grouping techniques. This combination allows us to analyze the potential of separating users in quasi-orthogonal groups, when the number of total Uniform Planar Array (UPA) elements at the Base Station (BS) is large, i.e., in massive MIMO, in order to improve performance and simplify complexity. Furthermore, IOMIM linear precoding techniques tend to offer varying throughput to different UEs depending on their received signal-to-noise ratio ($\mathrm {SNR}$); i.e., a varying quality-of-service (QoS) among UEs. This problem has not been studied in detail before, although it has the potential to support future wireless communication applications. Due to channel correlation and received power variation among different UEs, we propose assigning different subcarrier frequencies through Orthogonal Division Multiplexing (OFDM) to different UE groups. This method was developed in \cite{TK_EA_TCOM2}, where it is called Combined Frequency and Spatial Division Multiplexing (CFSDM),
and is critical to achieving a balanced QoS to all UEs in a cell. Furthermore, in order to improve the power efficiency in each CFSDM group, we also propose Optimized Per-Group Power Allocation (OPGPA), a new method that achieves significant power savings at the BS while simultaneously achieving equal QoS to all UEs in a group. We furthermore combine IOMIM with JSDM-FA, which imposes orthogonality between groups, but has not been studied in realistic deployments, e.g., a mmWave cell.

This paper addresses all these open issues while providing a comparison of various downlink precoding techniques. We first study different sub-grouping (SBG) forming techniques within each CFSDM subcarrier, in order to achieve high throughput with low complexity. These SBG techniques form the inner precoders of each group. To meet the goal of forming subgroups, we exploit the Virtual Channel Model Beam (VCMB), which was originally presented in \cite{TK_EA_TCOM2,ZF-PGP}. VCMBs are created by projecting the actual UE channels to a Discrete Fourier Transform (DFT) type basis that helps exploit the channel's spatial domain characteristics. We propose and analyze three techniques for SBG, including JSDM-FA together with a careful inter-sub-group interference analysis. We then compare the performance of two types of outer precoding: a) Zero-Forcing Precoding (ZFP) \cite{Swindle,Tse,Shamai}, and b) Zero-Forcing Per-Group Precoding (ZF-PGP). Due to high channel correlation between different UEs, ZFP gains evaporate rapidly and a Per-Group Precoding within Groups (PGP-WG) precoder performs much better as we demonstrate.

The contributions of this paper can be summarized as follows:
\begin{enumerate}
\item It employs a realistic mmWave communications model for the massive MIMO downlink, that includes random UE blocking, with the path-loss exponent and multipath-fading distribution dependent on the blocking state.
\item It presents a comprehensive approach to dividing UEs in CFSDM groups based on their spatial and power features, then subdividing groups in sub-groups by SBG in order to improve performance and lower the system complexity.
\item It presents results for three types of SBG, including a detailed inter-sub-group interference analysis in the case of JSDM-FA.
\item It shows that due to the debilitating impact of mmWave channels, the outer precoder in each group faces high channel correlation that leads to very poor performance of ZFP.
\item It demonstrates that in general JSDM-FA suffers a very high performance loss over Total Grouping (TG) and Simple Grouping (SG), due to its inherent significant inter-sub-group interference and high inter-UE channel correlation.
\item It shows the very high gains of ZF-PGP over ZFP for a wide range of $\mathrm {SNR}$.
\item It develops OPGPA, which is a new power-allocation strategy that allows for very high power efficiency.
\end{enumerate}
\underline{Notation:} We use small bold letters for vectors and capital bold letters for matrices. ${\mathbf A}^T$, ${\mathbf A}^H$, ${\mathbf A}^*$, ${\mathbf A}_{\cdot, i}$, ${\mathbf A}_{i,\cdot}$, and ${\mathbf A}_{k,l}$ denote the transpose, Hermitian, conjugate, complex conjugate, column $i$, row ${i}$, and row $k$, column $l$ element of matrix ${\mathbf A}$, respectively. Further, $\mathrm {tr}({\mathbf A})$ denotes the trace of a (square) matrix $\mathbf A$. ${\mathbf S}^T$ denotes a selection matrix, i.e., of size $k\times n$ with $k<n$ consists of rows equal to different unit row vectors ${\mathbf e}_i$ where the row vector element $i$ is equal to $1$ in the $i$th position and is equal to $0$ in all other positions, the specific ${\mathbf e}_i$ vectors used are defined by the desired selection.  ${\mathbf F}_{N}$ denotes the DFT matrix of order $N$, $\mathrm {diag}[x_1,\cdots, x_k]$ is the diagonal matrix with main diagonal equal to vector $[x_1,\cdots, x_k]^T$, and ${\mathbf I}_N$ denotes an identity matrix of dimension $N\times N$. We use ${\mathbf h}_{d,g,k,n}$ for the downlink channel of user $k$'s antenna $n$ in group $g$. ${\mathbf H}_g$ is the downlink channel of group $g$, while ${\tilde {\mathbf H}}_{g,v}$ is its projection to the Virtual Channel Model (VCM) basis.
\section{{S}ystem {M}odel and {P}roblem {S}tatement}
\subsection {{M}mWave {C}hannel {M}odel {E}mploying {R}andom UE {B}locking}
We assume a dense population of UEs that are uniformly distributed within a
cell \cite{MV2}. The BS UPA has height $h$ and employs an $x$ (horizontal) and $z$ (vertical) orientation\footnote{Any UPA or Uniform Linear Array (ULA) orientation would result in similar results and could be used with success in our model.}.
The total number of UE in the cell is $N_{UE}$. Each UE employs a single, uniformly radiating antenna. By employing Time Division Duplexing (TDD), the downlink channels will be reciprocal to the uplink ones. The channel between UE $n$ ($1\leq n\leq N_{UE}$) and the BS is denoted by ${\mathbf h}_n$. With $P=1$ multipath components \cite{TK_EA_TCOM2}, due to mmWave conditions \cite{MV1,MV2}, we get
\begin{equation}
{\mathbf h}_n=\tilde{g}_n\left({\mathbf a}_z(\theta_{n})\otimes{\mathbf a}_x(\theta_{n},\phi_{n}) \right),
\end{equation}
where $\otimes$ denotes Kronecker matrix product, $\tilde{g}_n =g_n\exp(j2\pi b_n)$ is the multipath fading complex coefficient of amplitude $g_n=|\tilde{g}_n|$ and phase $b_n$, uniformly distributed in $[0,2\pi]$, $\theta_{n}$, $\phi_{n}$ represent UE $n$'s ($1\leq n\leq N_{UE}$) elevation and azimuth angle, respectively,
        \begin{equation}
        \begin{split}
        {\mathbf a}_x(\theta_{n},\phi_{n}) = &[1, \exp(-j{2\pi}D\sin(\theta_{n})\cos(\phi_{n})),\cdots,  \exp(-j{2\pi} D(N_{u,x}-1)\sin(\theta_{n})\cos(\phi_{n}))]^T,\\&
         {\mathbf a}_z(\theta_{n}) = [1, \exp(-j{2\pi}D
         \cos(\theta_{n})),\cdots,  \exp(-j{2\pi D(N_{u,z}-1)}
         \cos(\theta_{n}))]^T,
         \end{split}
         \end{equation}
         with $D =\frac{d}{\lambda}$, $d$ being the distance between adjacent antenna elements, $\lambda$ the wavelength, and $N_{u,x},~N_{u,z}$ representing the number of elements of the UPA in the $\mathit{x}$ and $\mathit{z}$ direction, respectively. The total antenna elements at the BS equal to $N_T=N_{u,x}N_{u,z}$ (the number of rows in ${\mathbf h}_n$). The instantaneous received $\mathrm {SNR}$ at UE $n$ ($1\leq n\leq N_{UE}$) under the breakpoint model \cite{Molisch} is
         \begin{equation}
         \mathrm {SNR}_n =  g_n^2{\mathrm {SNR}_0}\left(\frac {R_{break}}{R_n}\right)^k,\label{eq_propag}
         \end{equation}
         for $R_n\geq R_{break}$, where $R_n$ is the distance between the UPA and the UE, $R_{break}$ is the break distance \cite{Molisch}, ${\mathrm {SNR}_0}=\frac{E_{s,0}}{N_0}$ is the $\mathrm {SNR}$ at $R_n=R_{break}$, where $E_{s,0}$ is the symbol energy and $N_0$ is the one-sided noise Power Spectral Density (PSD), $k$ is the path-loss exponent, and $ g_n^2$ is the power gain of the fading. As in \cite{MV1,MV2}, if UE $n$ ($1\leq n\leq N_{UE}$) is blocked, it is non-line-of-sight (NLOS) and we use $k=k_{NLOS}$ and $ g_n$ is Nakagami with $m=m_{NLOS}$, while when the UE is not blocked, it is line-of-sight (LOS) and we apply $k=k_{LOS}$ and $g_n$ is Nakagami with $m=m_{LOS}$, with $k_{NLOS}>k_{LOS}$ and $m_{NLOS}< m_{LOS}$. \\
         \subsection{Problem Statement}
From \cite{ZF-PGP}, an equivalent cell downlink channel receiving equation, after normalization and encompassing both large-scale and small-scale effects \cite{Molisch}, i.e., propagation loss and multipath fading, respectively, together with noise effects (including Additive White Gaussian Noise (AWGN) and Multiple-Access Interference (MAI)) can be written in the virtual domain.\footnote{This is the channel representation in the VCM basis, also called the beam-domain representation in the literature .} Toward this end, let's start by defining ${\mathbf y}_d$ to be the downlink received vector over all users and antennas of size $N_{UE}\times 1$, ${\mathbf G}$, and ${\mathbf x}_d$ to be the $N_{UE}\times 1$ vector of transmitted symbols\footnote{We assume that there is one symbol per receiving antenna in (\ref{final}), for simplicity.} drawn independently from a QAM constellation. Also define the unit $\mathrm{SNR}$ downlink virtual channel matrix to be ${\mathbf H}_{u,v}$, of size $ N_{UE}\times N_T$ downlink for all $N_{UE}$ UEs with its rows being the corresponding UE Hermitian of the uplink channel vector, employing $N_T$ receiving antennas at the BS, \cite{ZF-PGP,TK_EA_TCOM2}. Then, the downlink receiving equation is as follows
\begin{equation}
{{\mathbf y}_d} = \sqrt{\mathrm{SNR}_0}{{\mathbf H}}_{d,v}  {\mathbf G} {\mathbf x}_d + {{\mathbf n}}_{d,AWGN} + {{\mathbf n}}_{d,MAI} \label{final},
\end{equation}
where ${\mathbf n}_{d,AWGN}$ represents the complex circularly symmetric Gaussian noise of mean zero and variance per component $\sigma_{d}^2 = 1$ (after normalization by dividing the original receiving equation by the standard deviation of the AWGN noise), and ${\mathbf n}_{d,MAI}$ represents the multiple-access interference (MAI) between sub-groups, present only in the JSDM-FA case. We focus on the input-output mutual information $I({\mathbf x_d};{\mathbf y_d} )$ maximizing downlink precoding problem, where we assume that the channel is known at both the transmitter and the receiver(s)\footnote{In \cite{ZF-PGP} we show that estimated channels can be used successfully instead of the perfect channel knowledge assumed here.}, which can be cast as
\begin{eqnarray}
\begin{aligned}
& \underset{\mathbf G}{\text{maximize}}
& & I({\mathbf x_d};{\mathbf y_d})\\
& \text{subject to}
& &  \mathrm {tr}({\mathbf G} {\mathbf G}^H) = N_{UE}, \\ \label{eq_MMIMO_1}
\end{aligned}
\end{eqnarray}
where the constraint is due to keeping the total power transmitted from the BS to all downlink users equal to the total power without precoding. It is well-known that this problem is complexity-burdened and thus grouping UEs offers a solution to this \cite{JSDM1, JSDM2,HARVEST_1,ZF-PGP}. However, in this paper we aim at offering balanced throughput to UEs, thus additional methods are needed to achieve this goal, as described below. Furthermore, employing data symbols from a finite-alphabet constellation, e.g., QAM in (\ref{final}), makes the problem more realistic, but at the same time more complex \cite{Yongpeng}. For example, in order to solve the problem the mutual information $I({\mathbf x}_d;{\mathbf y}_d)$ needs to be computed and this can be performed using the Gauss-Hermite (GH) quadrature approximation. For the MIMO channel model presented in (\ref{final}), the GH approximation toward evaluating $I({\mathbf x}_d;{\mathbf y}_d)$  is presented in Lemma 1 of \cite{TK_EA_QAM} and is also described for completeness in Appendix A.
  \subsection{ UE Grouping and Sub-grouping  }
  {\bf a) Pre-selection of VCMBs}
  \newline
By projecting the uplink UPA response vector to the complete orthonormal basis ${\mathbf B}_{VCM}=\left({\mathbf F}_{N_{u,z}}\otimes {\mathbf F}_{N_{u,x}}\right)$ where ${\mathbf F}_{N}$ represents the Discrete Fourier Transform (DFT) matrix of size $N$, \cite{TK_EA_TCOM2} showed that with a large number of array elements, e.g., $N_T\sim 100$ and with equal elements per dimension $(N_{u,x}=N_{u,z})$, this projection achieves a sparse representation of the UE channels with only a few components from the columns of ${\mathbf B}=\left({\mathbf F}_{N_{u,z}}\otimes {\mathbf F}_{N_{u,x}}\right)$ needed. 
Since only a few columns (VCMBs) of the orthonormal matrix $\left({\mathbf F}_{N_{u,z}}\otimes {\mathbf F}_{N_{u,x}}\right)$ are needed to characterize each channel, a significant dimensionality reduction is available by employing the VCMBs. Furthermore, for spatially distant UEs, different users form quasi-orthogonal groups of non-intersecting VCMBs. In addition, VCMBs can be used to derive many useful spatial-domain features for the entirety of downlink channels in the cell. Here, we use the VCMB in the cell due to its spatial-feature-revealing capabilities. First, we extract the most ``loaded'' VCMBs in the cell, by determining the $N_{V,INIT}$ VCMBs that carry the most instantaneous power to UEs. The parameter $N_{V,INIT}$ is determined by the percentage of overall power in the cell we aim at capturing. Due to the nature of the VCMB structure, only a fraction of the total $N_T$ VCMBs are needed to guarantee that more than, e.g., $90\%$, of the total power is captured. This VCMB selection phase is called pre-selection (PS). The pseudo code for the PS algorithm is shown in Algorithm 1.
\newline

{\bf b) CFSDM-based Grouping for Balanced QoS}\newline
In mmWave communications, the existence of NLOS UEs which suffer a significant additional propagation power loss requires placing NLOS UEs in separate frequency sub-carriers, then employing higher power to improve the NLOS UE throughput, better balancing the throughput with that of the LOS UEs. Furthermore, due to high spatial correlation of UE channels, some UEs, although in the LOS class, will also experience low QoS. This QoS imbalance can be mitigated by adding another separate set of sub-carriers to accommodate these UEs, thus offering a solution to balancing the cell QoS. Thus, by employing a total of three or four sets of subcarriers in the form of CFSDM \cite{TK_EA_TCOM2}, depending on the number of UEs and the size of the cell, we can achieve a balanced QoS. Thus, there will be $N_G$ CFSDM groups in the cell, with $N_G=3~\text{or}~4$. The first group $G_1$ is the NLOS group formed by aggregating all the NLOS UEs, while the other groups, $G_2,\cdots, G_{N_G}$ denote LOS groups. The corresponding number of UEs in each CFDSM group is denoted by $N_{1},~N_{2},\cdots,~N_{N_G}$, respectively. The selection of UEs for each CFSDM group is based on maximum statistical decorrelation per group, i.e., each group presents relatively low correlation among its member UE channels.
\newline

{\bf c) SBG Techniques}
\newline
After CFSDM grouping of UEs, there is additional opportunity with SBG for improved performance or lower complexity. It is important to mention that sub-grouping employs the same frequency for all subgroups in a group, i.e., spatial multiplexing takes place to improve performance in each group.
After PS takes place, in the reduced dimension VCMB space comprising $N_{V,INIT}$ VCMBs, there are many alternatives one can use for further SG of different UEs in a group. In this paper we consider the following three:
\begin{enumerate}
\item
Employ all pre-selected VCMBs in a group, which is TG.
\item  Select only the strongest VCMBs in the group, which is SG, resulting in a final number of VCMBs per group, $N_{V,FINAL}=N_{N_g}$, $g=1,2,\cdots,N_{N_G}$, i.e., significantly smaller than $N_{V,INIT}$.
{\item JSDM-FA for semi-orthogonal sub-groups \cite{TK_EA_TCOM2, ZF-PGP}, which offers an $N_{V,FINAL}$ even smaller than SG, but in general it suffers intra-sub-group interference}. Due to the intra-group MAI issue, JSDM-FA can be applied by dividing each of the CFSDM groups in sub-groups. In other words, for each CFSDM group $G_k$, $k=1,~2,\cdots,~N_G$, we create $N_{S_k}$ sub-groups denoted as $G_{k,l}$ with $l=1,\cdots,N_{S_k}$, which apply JSDM-FA, where the number of subgroups per group, $N_{S_k}$, depends on the VCMB strength in each group. We use $N_{G_{k,l}}$ to denote the number of UEs in sub-group $G_{k,l}$, $k=1,2,\cdots,~N_{N_G}$, $l=1,\cdots,N_{S_k}$. The selection of the UE members of each sub-group is based on the first few strongest VCMBs.
\end{enumerate}
It is important to stress that during SG and JSDM-FA, VCMBs that are unused by some sub-groups are set off, i.e., no power is transmitted over these VCMBs by these sub-groups. If other subgroups employ these VCMBs, then the potential for MAI between subgroups arises. Thus, although in the SB scenario, there is no MAI between groups, in the JSDM-FA case, there might be some inter-sub-group MAI to the other subgroups that do have the affecting VCMBs unused is due to the lack of full orthogonality between JSDM-FA sub-groups. Thus, a careful calculation of the MAI between sub-groups in the JSDM-FA case is required. The details of this MAI calculation are presented in Appendix B.

\begin{algorithm}[h]
\caption{PS algorithm}\label{global}
\begin{algorithmic}[1]
\For {$i=1$ to $N_T$}
\State calculate power of VCMB $i$, $P_i=||H_{d,v}[:,i]||^2$;
\EndFor
\State sort in descending order the vector ${\mathbf P}=[P_1~P_2\cdots P_{N_T}]$, resulting in a new sorted power list vector ${\mathbf P}_s=[P_{s_1}~P_{s_2}\cdots P_{s_{N_{V,INIT}}}]$ and descending-order sorted VCMB list vector ${\mathbf v}_s$ 
\State select the first $N_{V,INIT}$ entries of ${\mathbf P}_s,~{\mathbf v}_s$ and denote them ${\mathbf P}_{PS},~{\mathbf v}_{PS}$, respectively and calculate the total power in ${\mathbf P}_{PS}$ as percentage of the original total power, i.e.,
$P_F=\frac{\sum_{i=1}^{N_{V,INIT}}P_{s_i}}{\sum_{i=1}^{N_T}P_i}$;
\State re-arrange the selected ${\mathbf v}_{PS}$ VCMBs per UE in descending channel power order, resulting in a $N_{UE}\times N_{V,INIT}$ matrix ${\mathbf M}_{V,PS}$ used in further processing
\end{algorithmic}
\end{algorithm}
%
\subsection {Efficient MU-MIMO Outer Precoding through ZF-PGP for Spatially Correlated Groups  }
After groups and sub-groups are selected and the pre-beamformer (inner precoder) is constructed, an MU-MIMO linear precoder (outer precoder) is deployed to offer individual and high data rate streams to each UE in a sub-group. References \cite{JSDM1,molisch_rapp} envisaged this type of hybrid precoding in order to achieve high throughput to each user in a group by employing a Zero-Forcing Precoder (ZFP). However, since all the UEs in a formed group possess spatial similarity, the channels within a group are highly correlated. Thus, ZFP results in low data rates and low spectral efficiency. Here we apply ZF-PGP \cite{ZF-PGP} which combines the benefits of ZFP and Per-Group Precoding within Groups (PGP-WG) \cite{TE_TWC} in order to improve the performance, as illustrated in Fig. 1. ZF-PGP also employs the Virtual Additional Antenna Concept (VAAC) \cite{ZF-PGP} that delivers two symbols to each UE, thus doubling the high $\mathrm{SNR}$ throughput of ZF-PGP over ZFP.
\begin{figure}[h]
\centering
\setcounter{figure}{0}
\vspace{-0 mm}
\includegraphics[height=3.6in,width=5.25in]{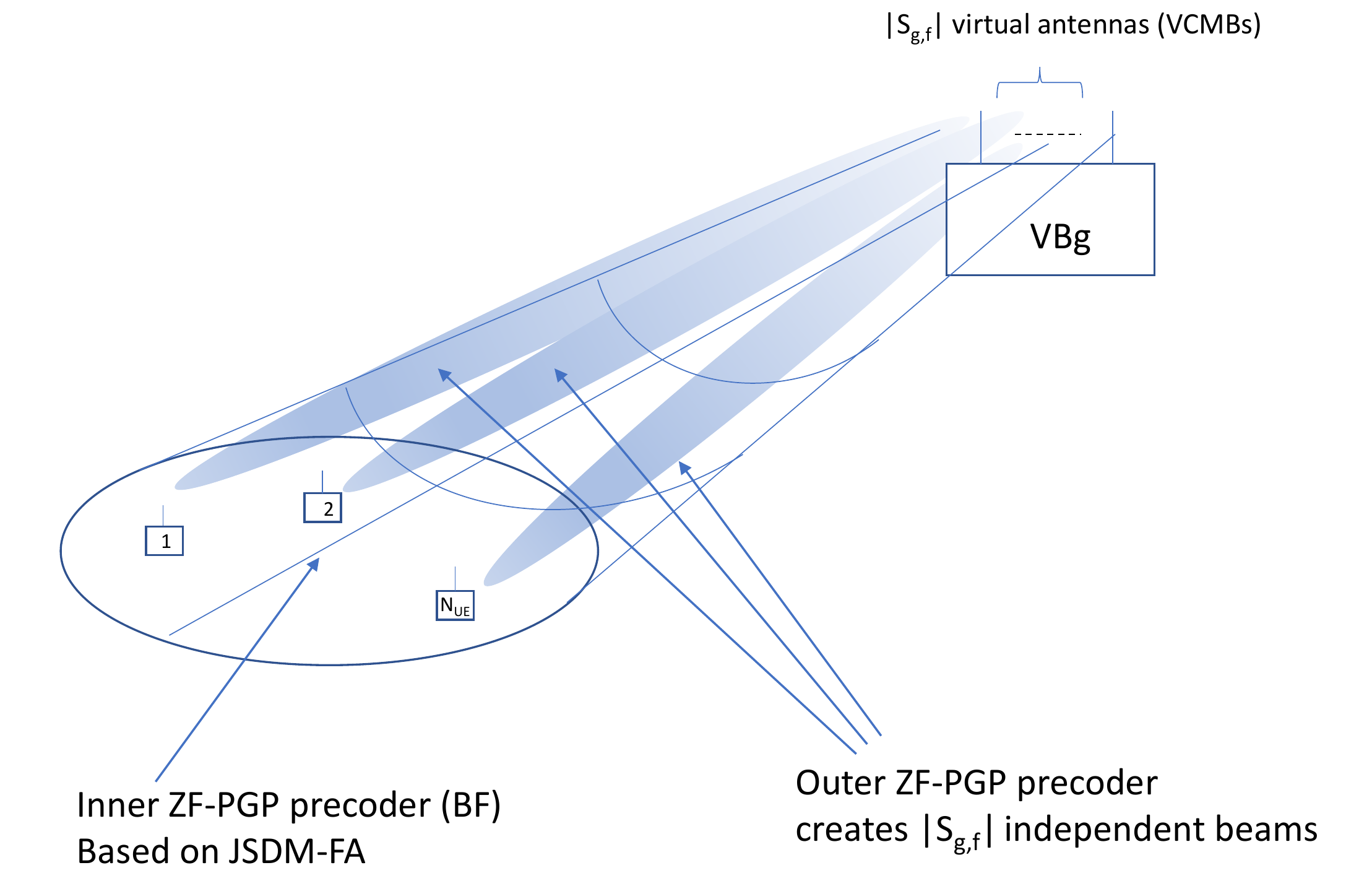}
	\caption{Outer group precoder. After inner precoding, the outer ZF-PGP linear precoder creates ${\cal S}_{g,f}$ independent data streams, one for each UE. ${\cal S}_{g,f}$ denotes the final number of VCMBs in the sub-group.}
\end{figure}

\subsection {Theoretical MU-MIMO Outer Precoding through VAAC-PGP for Comparison of Performance  }
When the cell radius grows, then there is additional power propagation loss incurred as well as additional correlation present to the channels of more distant UEs. Due to this, ZF-PGP can perform poorly for these UEs. A modified, improved, higher-performance precoder is studied for these UEs. VAAC-PGP is a PGP-WG \cite{TE_TWC} in conjunction with the application of the Virtual Additional Antenna Concept (VAAC)  and it can offer significant gains over ZF-PGP, albeit with higher complexity. Its higher complexity emanates from the fact that it requires knowledge of the overall group receiving vector and one of the downlink group channel's left singular vectors at each UE. Because of these mostly unrealistic demands, VAAC-PGP is employed only to give us some more fundamental understanding of the underpinnings of the channel correlation in the mmWave cell.

 Let LOS group $G_k$, with $k=2,~3,\cdots,~N_G$ employ this concept. The VAAC-PGP precoder adds one additional virtual antenna \cite{ZF-PGP} per UE, similar to the ZF-PGP one, but it does not include a ZF part, i.e., it does not offer the benefits of ZF precoding. This means that the UEs involved in VAAC-PGP need to know (through side information sent by the BS), or estimate the channel's right singular values. An additional benefit of VAAC-PGP, besides higher throughput to stressed UEs, is that one can rotate the use of the channel's right singular values and then by-reassigning data, a throughput-balancing effect takes place. In other words, all UEs involved in the VAAC-PGP precoding have the same throughput.

The VAAC-PGP is derived as follows. Assume that a group $g$ $(g=G_2,~G_3,~\cdots,~G_{N_G})$ applies VAAC-PGP. From \cite{JSDM1}, the equation for group $g$ is
\begin{equation}
{\tilde{\mathbf y}}_{g} = {\mathbf H}_{g,v} {\mathbf P}_{g} {\mathbf x}_{g} + {\mathbf n}_{g},
\end{equation}
where ${\mathbf H}_{g,v}$ is the VCM group's downlink matrix of size $N_{g}\times N_{g}$, ${\mathbf y}_{g}$ is the group's size $N_{g}$ reception vector, ${\mathbf P}_{g}$ is the $N_g\times 2\cdot N_g$ precoding matrix, ${\mathbf x}_{g}$ is the $2\cdot N_g\times 1$ data symbol vector, and ${\mathbf n}_{g}$ is the corresponding AWGN noise. For the data vector ${\mathbf x}_{g}$, we assume without loss of generality, that ${\mathbf x}_{g}=[x_{g,1,1}~ x_{g,1,2}~ \cdots x_{g,N_g,1}~ x_{g,N_g,2}]^T $, where $x_{g,i,k}$ with $i=1,2,\cdots,~ N_g$ and $k=1,2$ is the $i$th UE $k$th data symbol.

Then, the VAAC-PGP of the group solves the following optimization problem
\begin{eqnarray}
\begin{aligned}
& \underset{{\mathbf P}_g}{\text{maximize}}
& & I({\mathbf x}_g;{\mathbf y}_g)\\
& \text{subject to}
& &  \mathrm {tr}({\mathbf P}_g {\mathbf P}_g^H) = 2\cdot N_{UE}, \\ \label{eq_MMIMO_2}
\end{aligned}
\end{eqnarray}
under the constraint that ${\mathbf P}_g={\mathbf U}_{g,v}{\mathbf S}_{P}{\mathbf V}_{P}^H$, with ${\mathbf U}_{g,v}$ being the matrix of left singular vectors of ${\mathbf H}_{g,v}$, ${\mathbf S}_{P}$ being the singular value matrix of the VAAC-PGP precoder of size $N_g\times N_g$, and ${\mathbf V}_{P}$ of size $N_g\times 2\cdot N_g$ is the matrix of right singular vectors of ${\mathbf P}_g$. The constraint of (\ref{eq_MMIMO_2}) is placed in order to guarantee that the total transmitted power before and after recoding stays the same. An equivalent way of expressing (\ref{eq_MMIMO_2}) is through the singular value decomposition (SVD) of ${\mathbf P}_g={\mathbf U}_{g,v}{\mathbf S}_{P}{\mathbf V}_{P}^H$, as $\sum_{i=1}^{N_g}s_i^2=2\cdot N_g$, with $s_i$ being the $i$th singular value of ${\mathbf S}_{P}$. Due to applying PGP-WG, the matrix of the right singular vectors of the precoding matrix ${\mathbf V}_{P}$ observes a block-diagonal structure \cite{TE_TWC} as follows
\begin{equation}
\begin{split}
{\mathbf V}_{P}^H=\left[\begin{array}{cccccc}
{\bf v}_{p,1} & {\bf 0} & {\bf 0} & \cdots & {\bf 0} & {\bf 0}\\
{\bf 0} & {\bf v}_{p,2} & {\bf 0} & \cdots & {\bf 0} & {\bf 0}\\
{\bf 0} & {\bf 0} & {\bf v}_{p,3} & \cdots & {\bf 0} & {\bf 0}\\
\vdots & \vdots & \vdots & \ddots & \vdots & \vdots\\
{\bf 0} & {\bf 0} & {\bf 0} & \cdots &{\bf v}_{p,N_g-1} & {\bf 0}\\
{\bf 0} & {\bf 0} & {\bf 0} & \cdots &{\bf 0} & {\bf v}_{N_g}\\
\end{array}
\right],
 \end{split}
\end{equation}
where ${\bf v}_{p,i}$, $i=1, \cdots, N_g$ are $1\times 2$ row vectors of unit norm. As it is shown in Appendix C, the VAAC-PGP precoder has all its singular values equal to $\sqrt 2$, i.e. ${\mathbf S}_{P}=\sqrt 2{\mathbf I}_{N_g}$, due to the VAAC applied in the precoding process, i.e., the fact that the IOMIM optimal precoder with VAAC applied always needs all the possible power set to the useful antenna (see Appendix C). The UEs can decode their data by forming the inner product of the proper left singular vectors with the received vector of the group, ${\mathbf y}_g$.

%

 \subsection{Optimized Per-Group Power Allocation (OPGPA) for Equal QoS  and Reduced Average Power}
 Due to the nature of PGP and ZF-PGP, a previously uninvestigated possibility is available toward reducing the average power of a group's precoding, by allocating different powers to different effective singular values  employed in the ZF-PGP. We propose OPGPA in order to dramatically reduce the average power employed in a group, under medium channel correlation, to achieve equal throughput to all UEs in a group. OPGPA's premise is the basic idea that for a specific group-wide pre-set throughput goal of $I_S,$ with ($0\leq I_S\leq 2\mathrm{log}_2(M)$)\footnote{The reason for the $2$ in the equation is due to the VAAC concept that sends two symbols per receiving antenna. Please see Appendix B.} delivered to each UE in a group, based on ZF-PGP, the larger effective channel singular values defined in \cite{ZF-PGP} will need lower $\mathrm{SNR}$ than the smaller ones in order to achieve this goal. In addition, due to the PGP part of ZF-PGP, the power sent over one effective singular value, e.g., $s_{g,v,eff,m}\doteq \frac{1}{w_m}= \left({\sum_{m'=1}^{N_{d,g}}\frac{|({\mathbf V}_{g,v})_{m,m'}|^2}{s_{g,v,m'}^2}}\right)^{-1/2}$, where $m=1,2,\cdots, N_g$ \cite{ZF-PGP}. Assume that PGP group $m$ employs an $\mathrm{SNR}$ at $r=1~\mathrm{m}$ equal to $\mathrm{SNR}_0$, as per (\ref{final}), then an equivalent reception model with ZF-PGP for the $m$th PGP group incorporating the VAAC model \cite{ZF-PGP} is

\begin{eqnarray}
\begin{aligned}
\left[\begin{array}{c  } {y_{m,1}}\\
{y}_{m,a}\end{array} \right]&=&
\left[\begin{array}{c c } s_{g,v,eff,m}^{(\mathrm{SNR}_0)}\sqrt{2} &  0\\
 0 & 0\end{array} \right]
{\mathbf V}_{PGP,m}^H
\left[\begin{array}{c  }  x_{m,1}\\
 x_{m,a} \end{array} \right] +
\left[\begin{array}{c  } { n_{m,1}}\\
{ n_{m,a}}\end{array} \right], \label {eq_OPGPA}
\end{aligned}
\end{eqnarray}
where  $s_{g,v,eff,m}^{(\mathrm{SNR}_0)}$ is the $m$th channel's effective singular value under ZF-PGP for the value of $\mathrm{SNR}_0$ employed (see (\ref{final})), the vector of the noise has variance per component equal to 1, and ${\mathbf V}_{PGP,m}$ represents the $2\times 2$ PGP part right singular vector matrix of the ZF-PGP.

Under OPGPA, we allow weaker PGP groups to apply higher $\mathrm{SNR}$ in order to achieve higher throughput, equal to the set acceptable one, $I_S$. This means that each input to the ZF-PGP in (\ref{eq_OPGPA}) is employing an appropriate power gain factor denoted as $k_m$, thus achieving an overall $\mathrm{SNR}$ for the $m$th PGP group of $k_m\cdot \mathrm{SNR}_0$. This has no effect to the other PGP groups in the precoder, because in PGP-WG each group is orthogonal to the other groups by construction. When OPGPA applies, we call $\mathrm{SNR}_0$ as the initial $\mathrm{SNR}$ to avoid confusion. The corresponding receiving equation under  OPGPA is then
\begin{eqnarray}
\begin{aligned}
\left[\begin{array}{c  } {y_{m,1}}\\
{y}_{m,a}\end{array} \right]&=&
\left[\begin{array}{c c } s_{g,v,eff,m}^{(\mathrm{SNR}_0)}\sqrt{2} &  0\\
 0 & 0\end{array} \right]
{\mathbf V}_{PGP,m}^H{\sqrt{k_m}}
\left[\begin{array}{c  }  x_{m,1}\\
 x_{m,a} \end{array} \right] +
\left[\begin{array}{c  } { n_{m,1}}\\
{ n_{m,a}}\end{array} \right]. \label {eq_OPGPA}
\end{aligned}
\end{eqnarray}
It is worth mentioning that the total transmitted $\mathrm{SNR}$ at $R=R_{break}$ for the two symbols sent in each group is equal to $2k_m\cdot \mathrm{SNR}_0$ in OPGPA. Then,
the following theorem and corollaries hold, as proved in Appendix D.
\begin{theorem}
Under the model of ZF-PGP in (\ref{eq_OPGPA}), the problem of maximizing the input-output mutual information over the matrix ${\mathbf V}_{PGP}$ with $\mathrm{SNR}_0$ employed is equivalent to using the following reception model, then maximizing the input, output mutual information over ${\mathbf V}_{PGP}$
\begin{eqnarray}
\begin{aligned}
\left[\begin{array}{c  } {y}\\
{y}_a\end{array} \right]&=&
\left[\begin{array}{c c } 1 &  0\\
 0 & 0\end{array} \right]
{\mathbf V}_{PGP}^H
\left[\begin{array}{c  }  x\\
 x_a \end{array} \right] +
\left[\begin{array}{c  } { n_{1}^{'}}\\
{ n_{a}^{'}}\end{array} \right], \label {eq_OPGPA_1}
\end{aligned}
\end{eqnarray}
where the noise vector is complex, cyclically symmetric Gaussian with mean zero and covariance matrix equal to $\left(\sqrt{2{k_m}}s_{g,v,eff,m}^{(\mathrm{SNR}_0)}\right)^{-2}{\mathbf I}_2$. Thus, the effective channel $\mathrm{SNR}$ is equal to $\mathrm{SNR}_{eff,m}=\left(\sqrt{2k_m} s_{g,v,eff,m}^{(\mathrm{SNR}_0)} \right)^{2}$. Furthermore, in the reception model of (\ref{eq_OPGPA_1}), the IOMIM precoder is the same for all $m=1,2,\cdots, N_g$ under a constant $\mathrm{SNR}_{eff,m}$.
\end{theorem}
 \setcounter{theorem}{0}
 \begin{corollary}
   The effect of multiplying the $m$ input vector by ${\sqrt k_m}$ is to increase the effective channel reception $\mathrm{SNR}$ by $k_m$.
   \end{corollary}
   \begin{corollary}
   The IOMIM precoder of (\ref{eq_OPGPA}) is a function of $\mathrm{SNR}_{eff,m}$, only. Therefore, for the same IOMIM-achievable $I_S$ value, the condition $\mathrm{SNR}_{eff,m}=\mathrm{SNR}_{req}(I_S)$ needs to be valid for all $m=1,2,\cdots,N_g$, where $\mathrm{SNR}_{req}(I_S)$ represents the effective $\mathrm{SNR}$ needed for achieving $I_S$ by the IOMIM PGP precoder.
   \end{corollary}
   \begin{corollary}
  For a QoS pre-set $I_S$, each PGP group can attain $I_S$, if the required $k_m$ in the model of (\ref{eq_OPGPA_1}) is set by $\sqrt{k_m} = {\sqrt \frac{\mathrm{SNR}_{req}(I_S)}{2}}\frac{1}{s_{g,v,eff,m}^{\mathrm{(\mathrm{SNR}_0)}}}$.
   \end{corollary}
   We would like to stress that due to Corollary 2, since all sub-group IOMIM precoders under OPGPA are the same, the BS needs to determine a single IOMIM precoder for the pre-set $I_S$, resulting in a major simplification in system design. Furthermore, with OPGPA the system achieves the same QoS for all UEs in a group. Finally, OPGPA achieves very high gains in power efficiency. Thus, OPGPA represents a major improvement in downlink precoding for future wireless networks.
Note that once the value of $I_S$ is set, one needs to only determine the corresponding value of $\mathrm{SNR}_{req}(I_S)$, then set the value of the gains $\sqrt{k_m}$ by $\sqrt{k_m} = {\sqrt \frac{\mathrm{SNR}_{req}(I_S)}{2}}\frac{1}{s_{g,v,eff,m}^{{(\mathrm{SNR}_0)}}}$, for $m=1,2,\cdots,N_g$. Then, the corresponding average $\mathrm{SNR}$ employed by the OPGPA system will be
\begin{equation}
{\mathrm{SNR}_{OPGPA}}= \frac{\mathrm{SNR}_0\cdot\mathrm{SNR}_{req}(I_S)}{N_g}\sum_{m=1}^{N_g} \frac{1}{{(s_{g,v,eff,m}^{(\mathrm{SNR}_0)})}^2},
\end{equation}
while the corresponding average required $\mathrm{SNR}$ without OPGPA will be
\begin{equation}
{\mathrm{SNR}_{NOPGPA}}= \mathrm{SNR}_0\cdot\mathrm{SNR}_{req}(I_S) \left(\frac {1}{{\min_{{m'}=1,\cdots,N_g}}\left\{{s_{g,v,eff,{m'}}^{(\mathrm{SNR}_0)}}\right\}}\right)^2,
\end{equation}
which shows that because of the high channel correlation, the approach without OPGPA will require much higher average $\mathrm {SNR}$ in order to achieve the requirement of all UEs meeting the set $I_S$ requirement. Also, a more prudent approach can be to set $\mathrm{SNR}_0$ as the minimum $\mathrm{SNR}$ and also set a maximum $\mathrm{SNR}$ equal to $\mathrm{SNR}_1$, where $\mathrm{SNR}_1>\mathrm{SNR}_0$. Then, depending on the scenario, some $I_S$ might not be possible to be met, even without OPGPA. Finally, notice that when $I_S<2\cdot\log_{2}(M)$, without OPGPA some UEs could experience much higher throughput than $I_S$, i.e., the group experiences non-balanced performance among the different UEs, but under OPGPA the performance and thus the QoS is equal among all the UEs in the group, provided that ${{\min_{{m'}=1,\cdots,N_g}}\left\{{s_{g,v,eff,{m'}}^{(\mathrm{SNR}_0)}}\right\}}\geq \sqrt{\frac{\mathrm{SNR}_0\cdot \mathrm{SNR}_{req}(I_S)}{\mathrm{SNR}_1}},$ as it can be seen easily.

 \section{Numerical Results}
In this section, we present our spectral efficiency results based on achievable input-output mutual information for an annular mmWave cell with internal radius $r_i$ and external radius $r_o$. The BS antenna array is at a height $h=3~\mathrm{m}$. Two cases are considered: a) $N_{UE}=10$, $r_i=1~\mathrm{m}$, $r_0=5~\mathrm{m}$, and b) $N_{UE}=20$, $r_i=1~\mathrm{m}$, $r_0=20~\mathrm{m}$. For the NLOS UEs we use $k_{NLOS}=4$ and $m_{NLOS}=2$ (Nakagami fading), while for the LOS UEs we use $k_{LOS}=2$ and $m_{LOS}=4$ (Nakagami fading). We employ an
annular cell of interior radius $r_i$ and exterior radius $r_o$ \cite{MV2}. For both cases, we consider multiple scenarios, including Total Grouping ZF-PGP (TGZF-PGP), Sub-group ZF-PGP (SGZF-PGP), JSDM-FA ZF-PGP, and corresponding results for Zero-Forcing Precoding (ZFP). In order to offer high QoS to the NLOS UEs, we employ CFSDM \cite{TK_EA_TCOM2}, so that the LOS UEs employ a different subcarrier frequency than the NLOS ones. Furthermore, the LOS UEs employ $\mathrm{SNR}_0=20~\mathrm{dB}$, while the NLOS UEs employ a 13 dB higher $\mathrm{SNR}_0$ than the LOS UEs, in order to balance the QoS between the NLOS group and LOS groups. Transmitted symbols are drawn from an $M=16$ QAM constellation with two LOS groups selected from the ${\mathbf S}_{EFF}$ matrix of the virtual LOS channel, after the pre-selection VCMB phase. The BS antenna has $N_T=100$, with equal elements in the $x$ and $z$ dimensions. We use the $\mathrm{SNR}_0$ at distance $r_i=1~\mathrm{m}$ from the BS array base. Also, in order to stress the differences in precoding between finite alphabet inputs and Gaussian ones, we present results for Gaussian inputs in some cases in addition to the QAM ones.
\subsection{Results with TG or SG without OPGPA}
For $N_{UE}=10$, $r_i=1~\mathrm{m}$, $r_0=5~\mathrm{m}$, i.e., a short-range cell deployment, with TG and $N_{V,INIT}=20$ ($98~\%$ of total cell power captured), we get $2$ NLOS users and the rest of the UEs are put in one group, $G_2$, initially. NLOS $G_1$ results are depicted in Fig. 2. We observe that due to employing CFSDM with higher power, the NLOS group is able to attain high throughput. In addition, we see that the Gaussian input performance is close to the finite-alphabet one until the finite alphabet reaches saturation. The corresponding TG results for LOS $G_2$ are shown in Fig. 3. We see that due to the high correlation in $G_2$, there is a large difference in performance between VAAC-PGP and ZF-PGP.
\begin{figure}
\centering
\includegraphics[height=3.6in,width=5.25in]{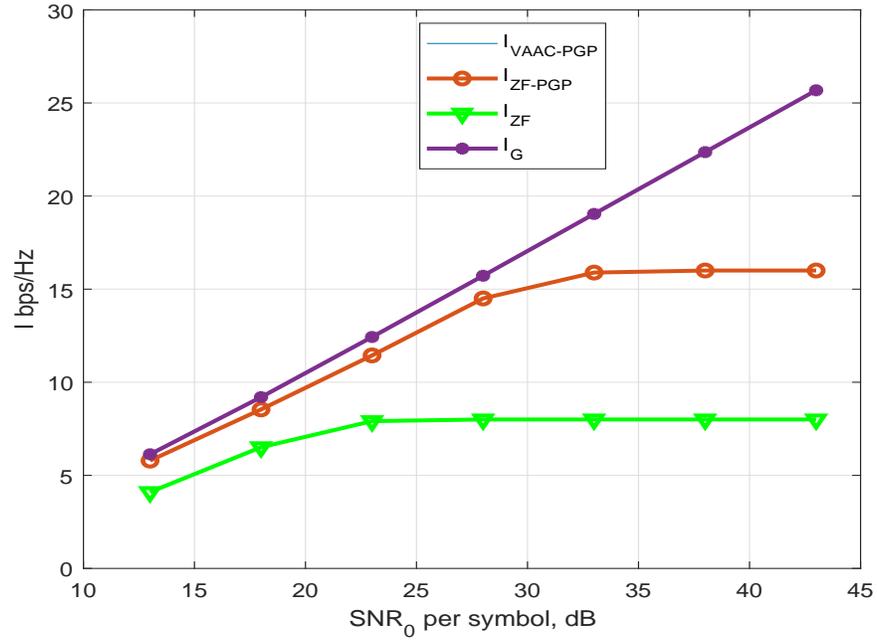}
\caption{Spectral efficiency for NLOS Group $G_1$ with TG employing 20 VCMBs for the first deployment scenario.}
\end{figure}

\begin{figure}
\centering
\includegraphics[height=3.6in,width=5.25in]{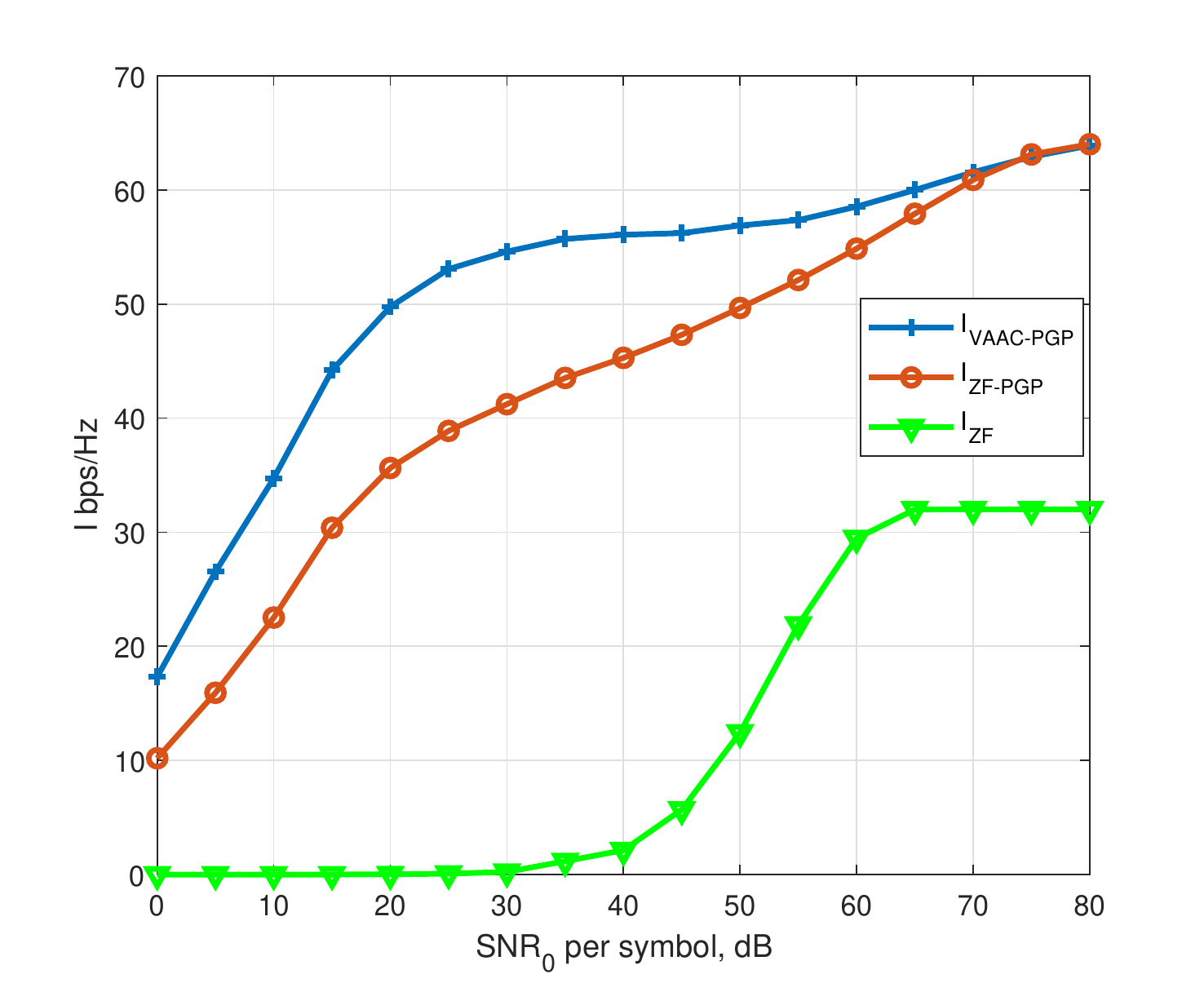}
\caption{Spectral efficiency for LOS UEs group $G_2$ for the first deployment scenario TG (20 VCMBs).}
\end{figure}
Next, we show results for $G_2$ with SG. We split $G_2$ in two frequency groups, based on similarities in their VCMBs, by allocating maximally distant spatial signatures in each group. Each of the two new groups $G_2$ and $G_3$ are allocated 4 VCMBs, then we apply SG with $N_{V,FINAL}=4$ in both groups. The results for $G_2$ and $G_3$ are shown in Fig. 4 and 5, respectively. We observe that because $G_2$ and $G_3$ are structured with minimum correlation, the difference in performance between VAACP and ZF-PGP diminishes.
\begin{figure}
\centering
\includegraphics[height=3.6in,width=5.25in]{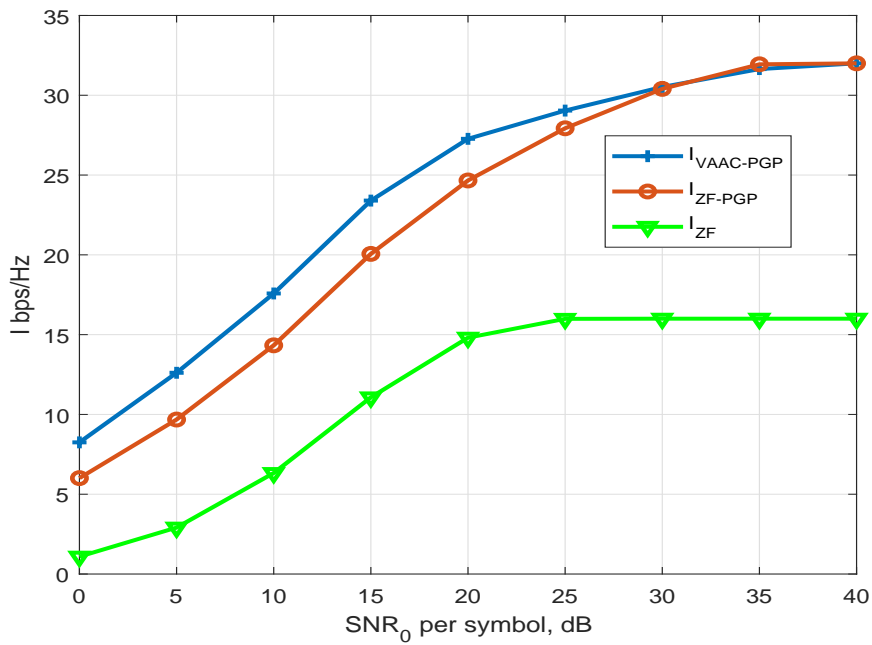}
\caption{Spectral efficiency for LOS Group $G_2$ for the first deployment scenario with SG, using 4 VCMBs.}
\end{figure}

\begin{figure}
\centering
\vspace{-2mm}
\includegraphics[height=3.6in,width=5.25in]{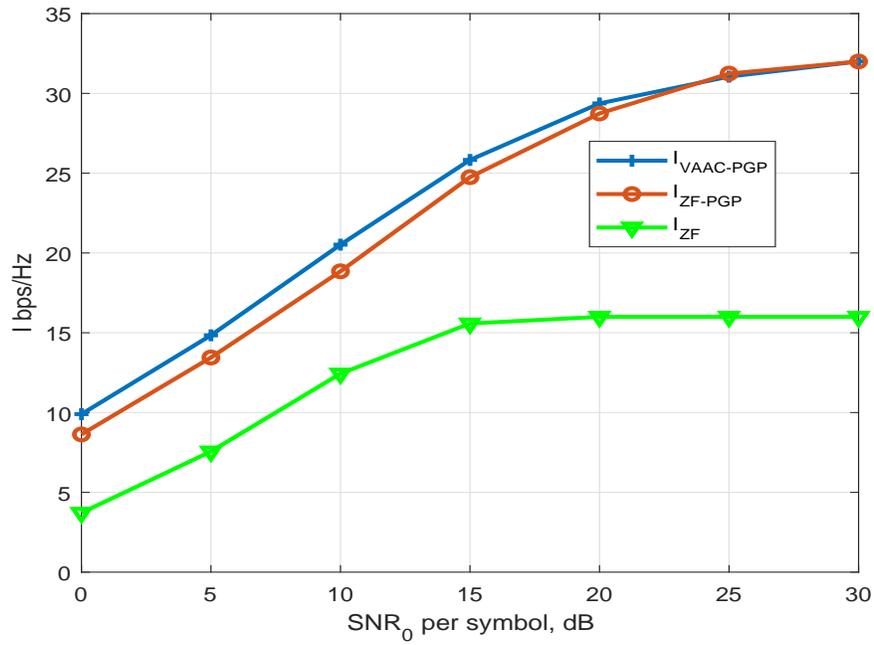}
\caption{Spectral efficiency for LOS Group $G_3$ for the first deployment scenario with SG, using 4 VCMBs.}
\end{figure}

For the second deployment scenario, $N_{UE}=20$, $r_i=1~\mathrm{m}$, $r_0=20~\mathrm{m}$, i.e., a longer range cell is considered.
For NLOS group $G_1$ with TG, we get the results shown in Fig. 6. We observe the same behavior as in the first deployment scenario.
\begin{figure}
\centering
\includegraphics[height=3.6in,width=5.25in]{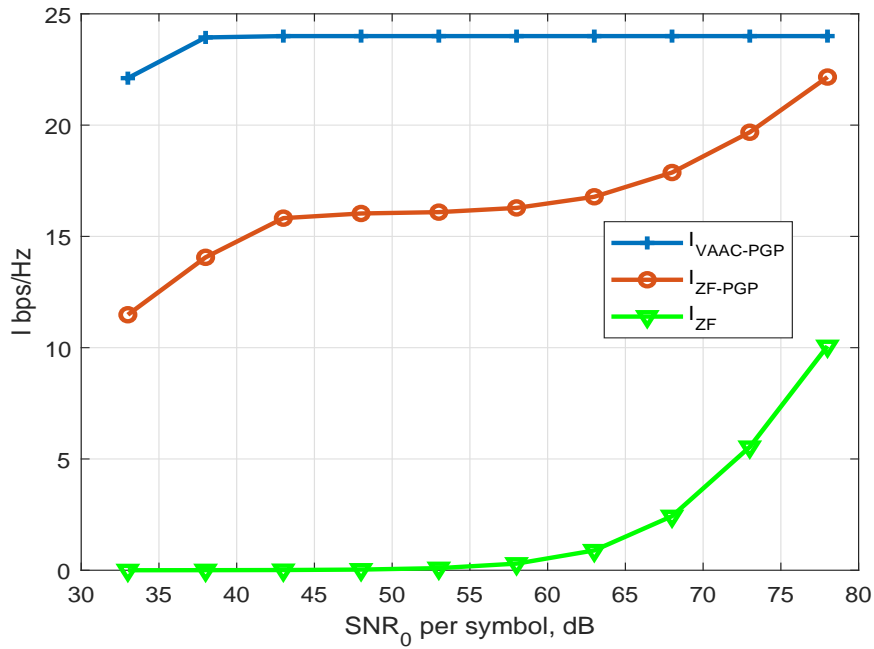}
\caption{Spectral efficiency for NLOS Group $G_1$ for the second deployment scenario with TG, using 10 VCMBs.}
\end{figure}
For LOS group $G_2$ with TG, we get the results shown in Fig. 7.
\begin{figure}
\centering
\includegraphics[height=3.6in,width=5.25in]{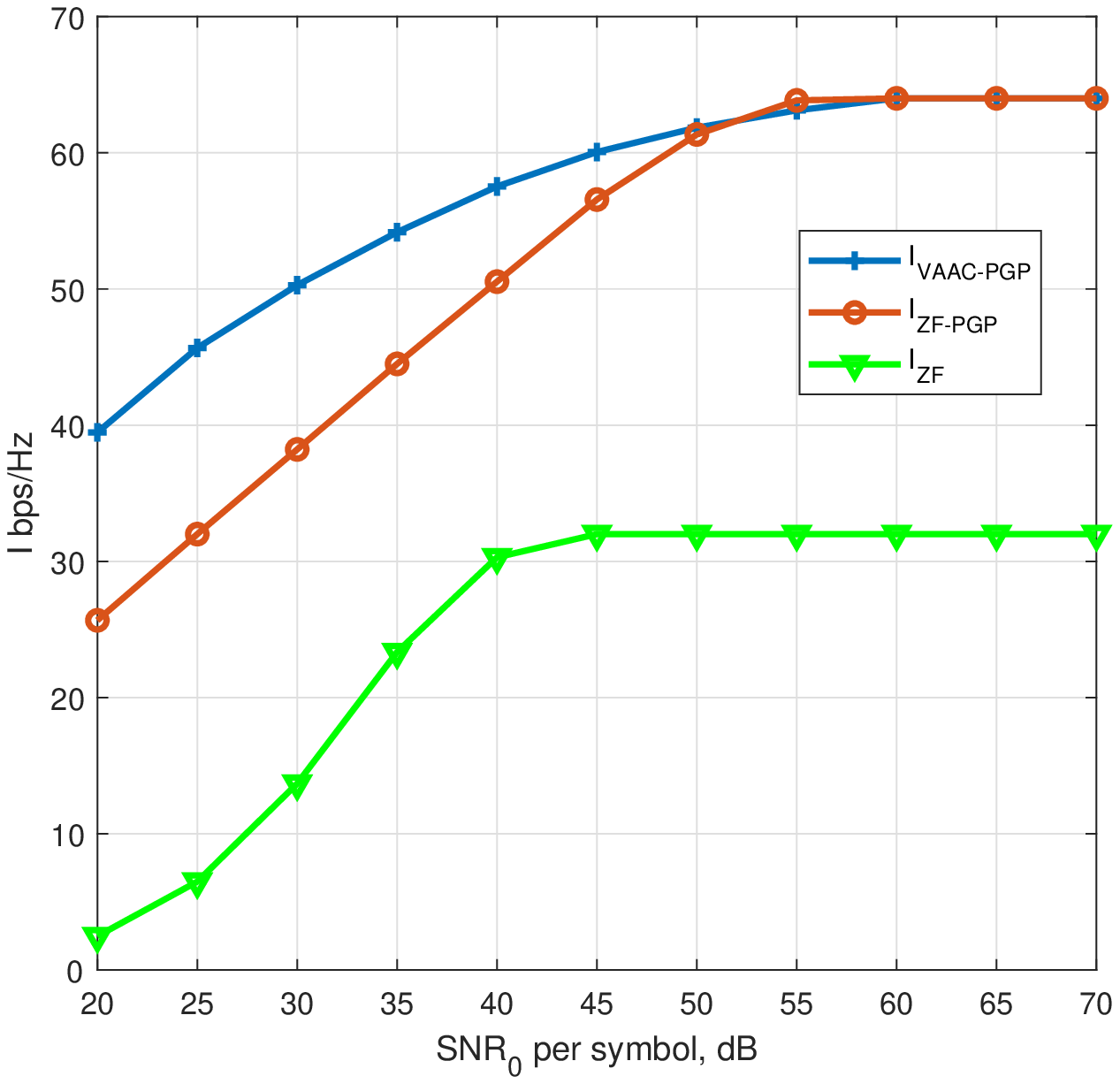}
\caption{Spectral efficiency for LOS Groups $G_2$ for the second deployment scenario with TG, using 20 VCMBs.}
\end{figure}
Next, we split the LOS UEs in two (distant) groups with 8 UEs each, and apply TG or SG. Fig. 7 shows results for $G_2$ in conjunction with TG. Since using $N_{V,FINAL}=20$ might be unrealistic, in Fig. 8, we show results with SG using $N_{V,FINAL}=8$.
\begin{figure}
\centering
\includegraphics[height=3.6in,width=5.25in]{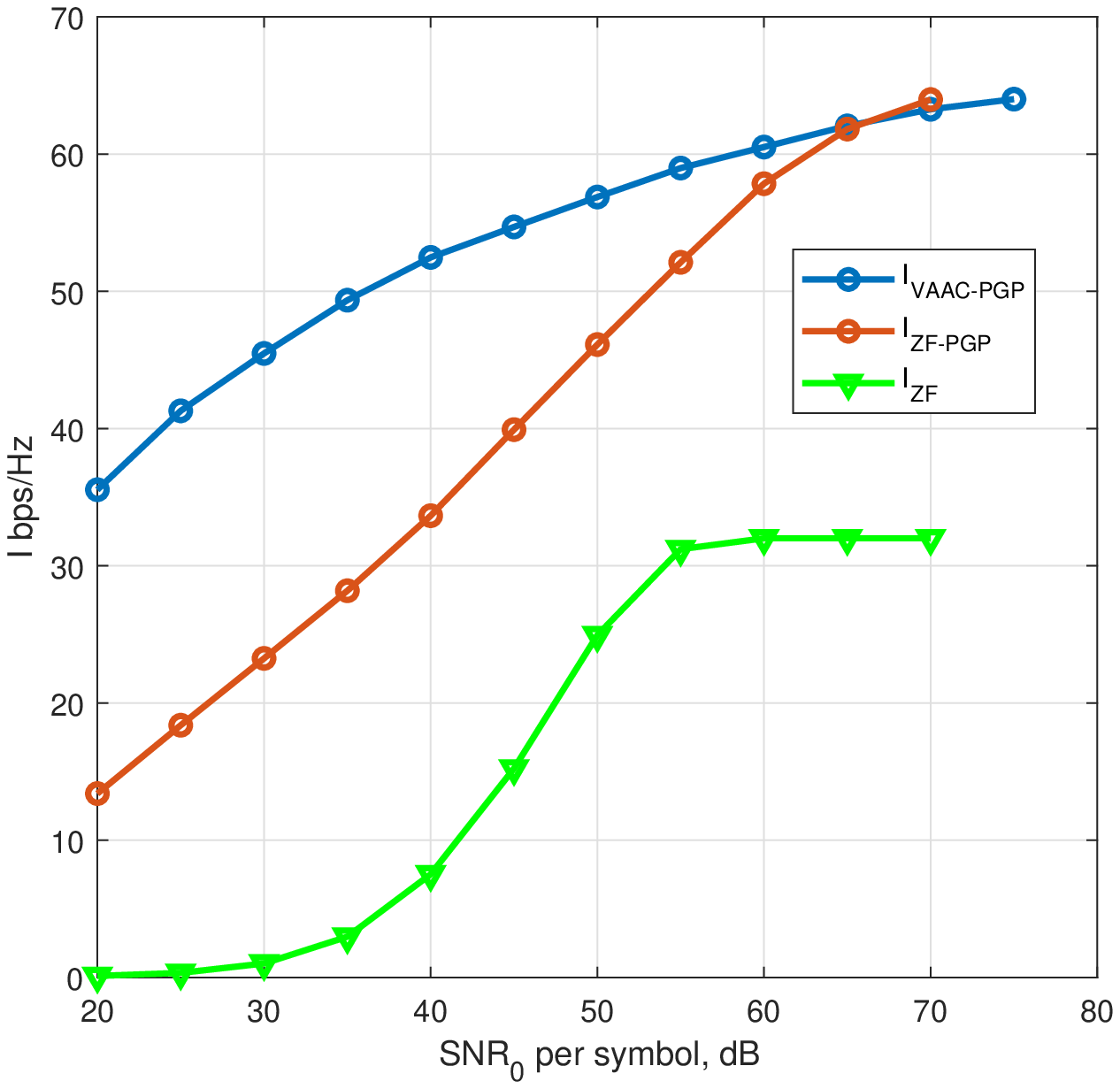}
\caption{Spectral efficiency for LOS Group $G_2$ for the second deployment scenario with SG, using 8 VCMBs.}
\end{figure}
Corresponding results for $G_3$ are shown in Fig. 9, both SG and TG achieve almost identical results. We see that $G_3$ exhibits high correlation, thus its performance is really low. In order to remedy the situation, further splitting of $G_3$ in two frequency groups is required, resulting in two new sub-groups of 4 UEs each, $G_3,~G_4$.
\begin{figure}
\centering
\includegraphics[height=3.6in,width=5.25in]{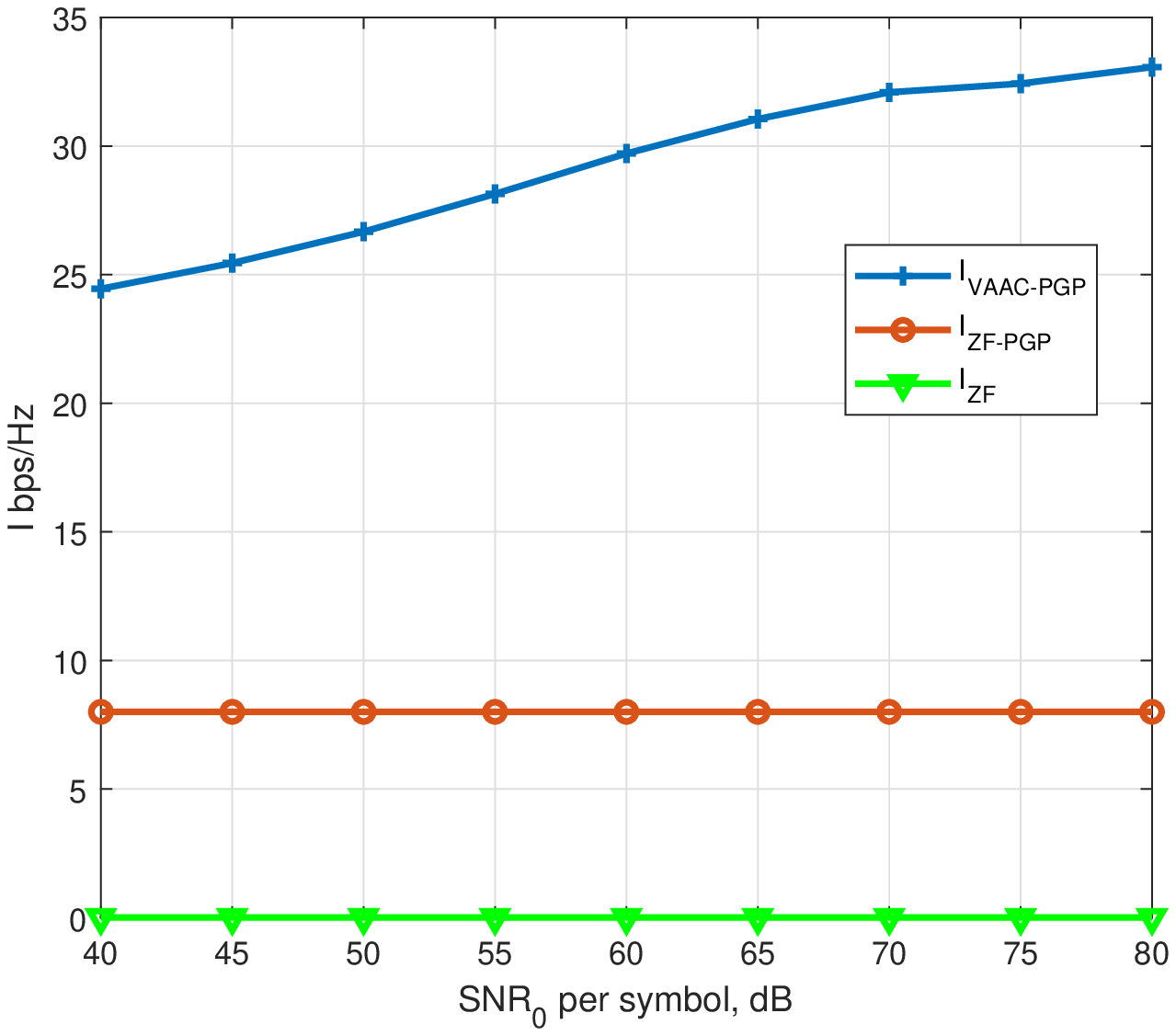}
\caption{Spectral efficiency for LOS Group $G_3$ for the second deployment scenario with TG, using 20 VCMBs (TG) or 8 VCMBs (SG), both having the same performance.}
\end{figure}
In Fig. 10 and 11 we show $G_3$ and $G_4$ results with $N_{V,FINAL}=4$.
\begin{figure}
\centering
\includegraphics[height=3.6in,width=5.25in]{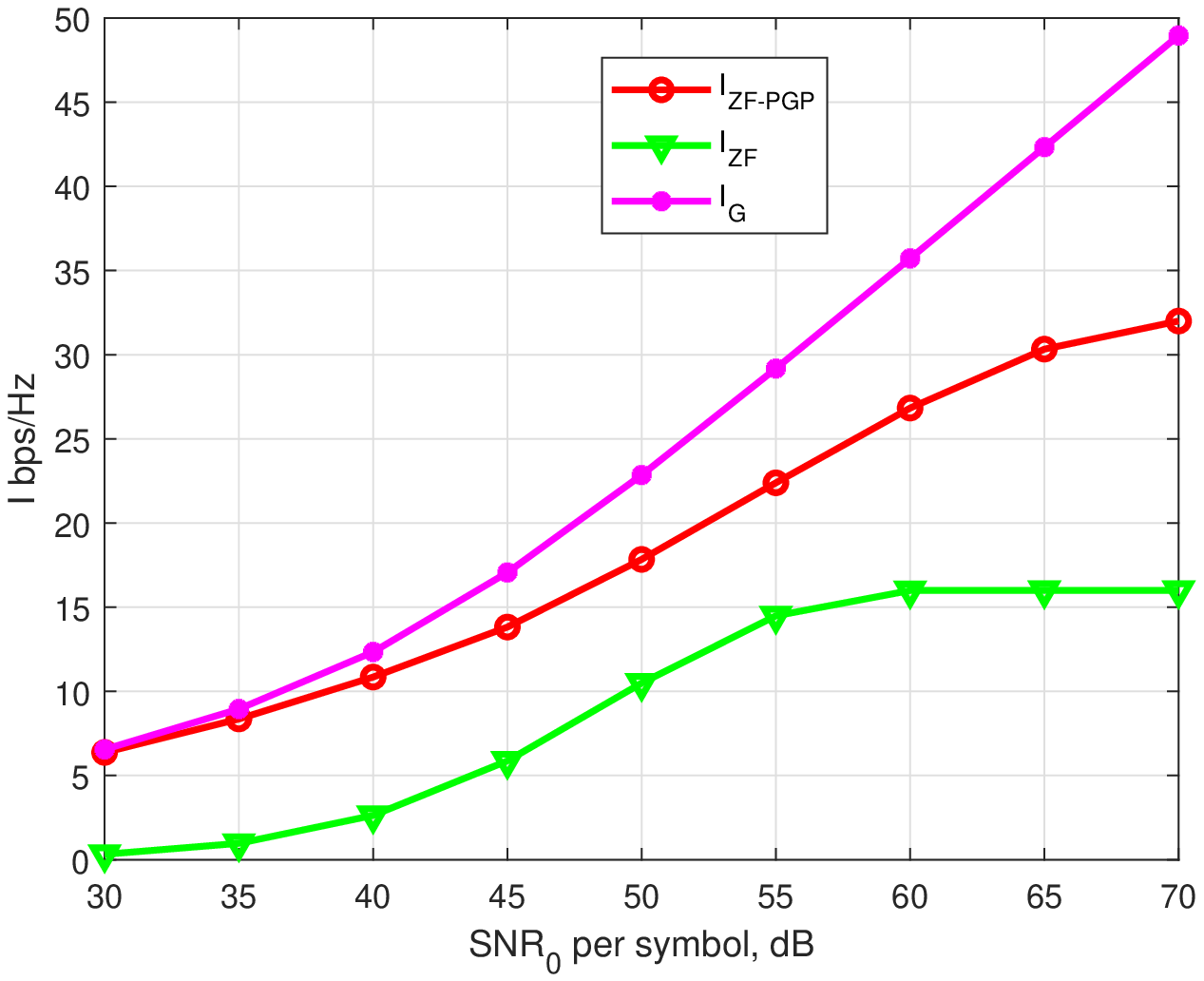}
\caption{Spectral efficiency for LOS Group $G_3$ for the second deployment scenario with SG, using 4 VCMBs.}
\end{figure}

\begin{figure}
\centering
\includegraphics[height=3.6in,width=5.25in]{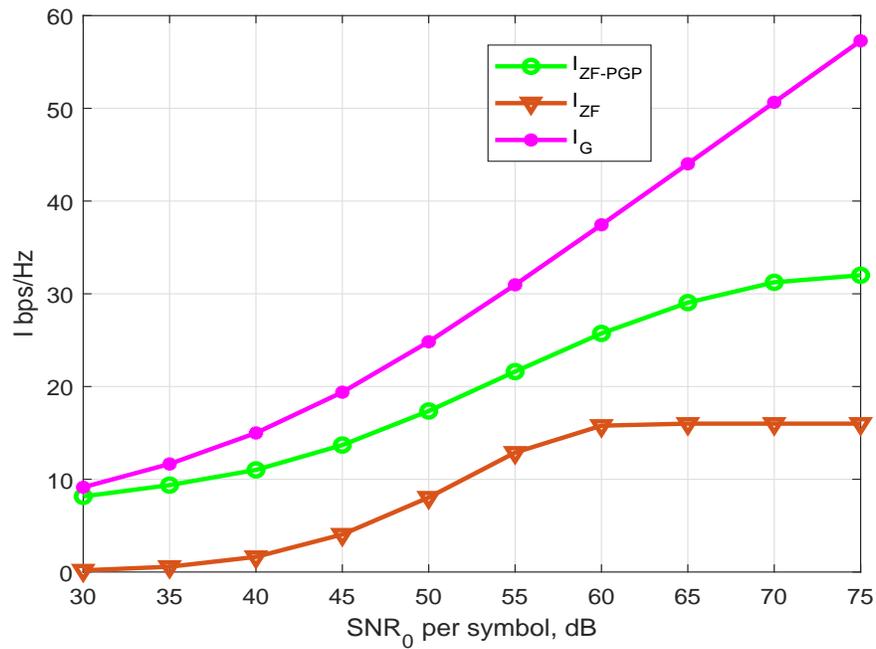}
\caption{Spectral efficiency for LOS Group $G_4$ for the second deployment scenario with SG, using 4 VCMBs.}
\end{figure}
For the Gaussian inputs, we observe in this case that their performance deviates significantly, even at lower $\mathrm{SNR}$, due to the additional correlation present in the current scenario. This is a quite important observation in this regard.
It is worth noting that the spectral efficiency achieved for the first scenario employing $\mathrm{SNR}_0=30~\mathrm{dB}$ for all three subcarriers in the SG case, is $\mathrm{SE}=26.33~\mathrm{bps/Hz}$, or SE per unit area, $\mathrm{SEUA}=0.3647~\mathrm{bps}/\mathrm{Hz}/\mathrm{m}^2$. For the second scenario, the corresponding numbers are at $\mathrm{SNR}_0=60~\mathrm{dB}$, $\mathrm{SE}=31.50~\mathrm{bps}/\mathrm{Hz}$, and $\mathrm{SEUA}=0.0252~\mathrm{bps/Hz/m^2}$, respectively.
\subsection{Results for TG or SG with OPGPA}
Here we present results with TG or SG employing the OPGPA concept, in order to offer equal QoS under lower average power consumption by the BS.
We focus on the second scenario and we apply OPGPA and NOPGPA with $\mathrm{SNR}_0=20~\mathrm{dB}$, in the case of groups $G_3$ and $G_4$ presented above, with four UEs each and SG. In Fig. 12 we present the corresponding results with respect to the pre-set QoS, $I_S$. For both cases, we see that by applying OPGPA, the system achieves more than 15 dB power reduction in the average $\mathrm{SNR}_0$ required. This is a very big improvement in the power requirement, while simultaneously OPGPA offers equal throughput to all UEs, i.e., it mitigates the imbalance in the QoS of different UEs.
\begin{figure}
\centering
\includegraphics[height=3.6in,width=5.25in]{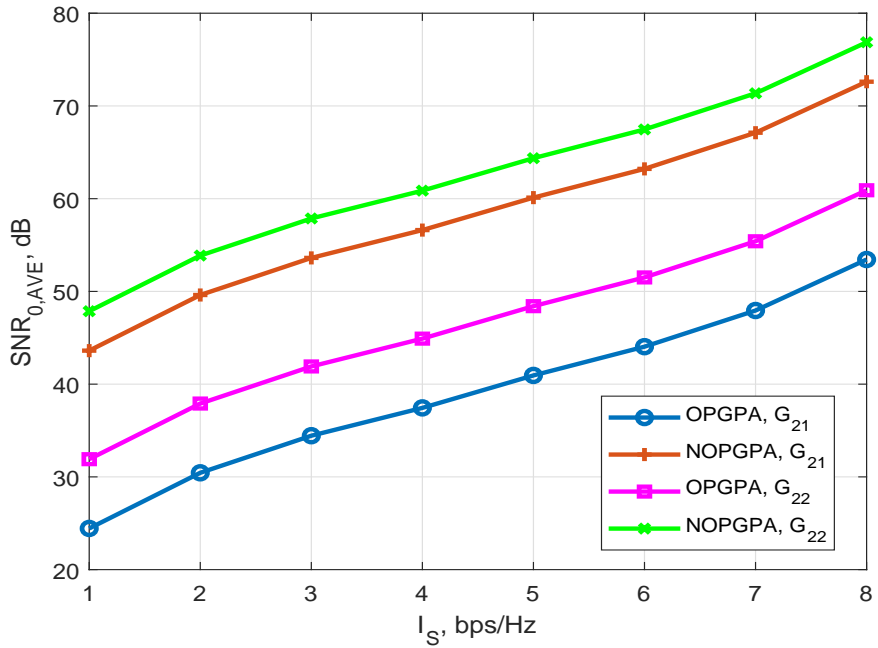}
\caption{OPGPA and NOPGPA $\mathrm{SNR}_{0,AVE}$ for LOS Group $G_3$ and $G_4$ for the second deployment scenario with SG, using 4 VCMBs.}
\end{figure}

\subsection{Results for JSDM-FA}
For JSDM-FA, we focus on the first scenario without OPGPA. In general, since full orthogonality between JSDMA-FA sub-groups is not possible, there is always MAI entering each UE's receiver, reducing its effective $\mathrm{SNR}$. The effective $\mathrm{SNR}$ for UE $i$ of sub-group $l$ of group $k$, $G_{k,l}$, $SNR_{eff,{ G}_{k,l},i}=\left( \frac{1}{\mathrm{SNR}_0}+\frac{1}{\sum_{l' \in { G}_{kl'}, l'\neq l}{\mathbf p}_{MAI,k,l,l'}(i)} \right)^{-1}$, where ${{\mathbf p}_{MAI,k,l,l'}(i)}$ is the MAI power to the $i$th UE in $G_{k,l}$ from ${ G}_{k,l'}$, evaluated in Appendix B for ZFP and ZF-PGP.

In Fig. 13 we present results for $G_{2,1}$ and $G_{2,2}$, comprising 3 and 5 UEs, respectively. We see that the performance of both sub-groups becomes MAI limited and it is significantly lower than the one achievable by, e.g., SG ZF-PGP. In addition, due to the high correlation in the sub-group UE channels, the performance of ZFP is negligible. For other scenarios with more UEs or longer cell range, the effects of correlation and MAI create an even harsher environment for JSDM-FA.
\begin{figure}
\centering
\includegraphics[height=3.6in,width=5.25in]{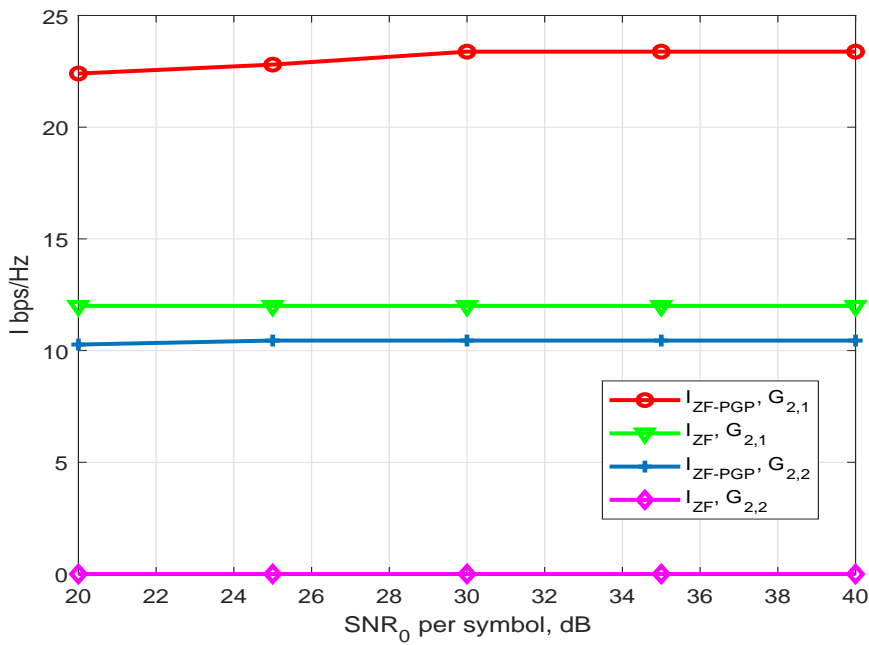}
\caption{JSDM-FA spectral efficiency for LOS Groups $G_{2,1}$ and $G_{2,2}$ for the first deployment scenario with JSDM-FA.}
\end{figure}

\section{Conclusions}
We have applied the concept of ZF-PGP in a mmWave massive MIMO cell, carefully modeling all channel intricacies, including UE blocking, path propagation, and multipath fading. It is shown that VCMBs offer many advantages in simplifying the channel representation, analyzing the system, and improving group and UE throughput. For the scenarios presented, it is shown that ZF-PGP offers significant throughput improvements, especially when the channel presents a medium-to-high degree of correlation. When such spatial correlation is present, ZF-PGP offers a $100\%$ improvement in throughput over ZFP. When even higher spatial correlation is present or the cell range becomes longer, ZF-PGP can offer more than $300\%$ throughput improvement over ZFP, albeit in the lower $\mathrm{SNR}$ region. We compare three different VCMB-based group-forming techniques which offer different advantages, depending on the application scenario. Finally, our work has demonstrated that when $N_{UE}<30$, only four subcarrier frequencies suffice for CFSDM to be used in the cell, in order to guarantee a balanced QoS for all UEs. We also introduced a new method, OPGPA, that significantly reduces the average  power transmitted per group, while it meets a specific QoS requirement for all UEs in a group. Finally, from our presented results, it becomes more evident that under the correlated channel conditions in a mmWave cell, the widely used Gaussian approximation falls short in accurately predicting the optimal precoder performance.
\bibliographystyle{IEEEtran}

\appendices

\section{GH Quadrature Approximation in MIMO Input-Output Mutual Information}
Let us consider a generic $N_t$ transmit antenna, $N_r$ receive antenna MIMO model as described by the following equation
\begin{eqnarray}
{\mathbf y} = {\mathbf H}{\mathbf G}{\mathbf x}+{\mathbf n}, \label{eq_1}
\end{eqnarray}
where ${\mathbf y}$ is the $N_r\times 1$ received vector, ${\mathbf H}$ is the $N_r\times N_t$ MIMO channel matrix,
 ${\mathbf G}$ is the precoder matrix of size $N_t\times N_t$, ${\mathbf x}$ is the $N_t\times 1$ data vector with independent components
each of which is drawn from the QAM
constellation of size $M$, ${\mathbf n}$ represents the circularly symmetric AWGN vector of size $N_r\times 1$, with mean zero and
covariance matrix ${\mathbf K}_n = \sigma^2_n {\mathbf I}_{N_r}$, where ${\mathbf I}_{N_r}$ is the $N_r\times N_r$ identity matrix, and $\sigma^2 =\frac{1}{\mathrm{SNR}}$.
$I({\mathbf x};{\mathbf y}) = H({\mathbf x})-H({\mathbf x}|{\mathbf y})= N_t\log_2(M)-H({\mathbf x}|{\mathbf y})$, where the conditional entropy, $H({\mathbf x}|{\mathbf y})$ can be written as \cite{Xiao2}
\begin{equation}\small
\begin{split}
H({\mathbf x}|{\mathbf y})  &= \frac{N_r}{\log(2)} +\frac{1}{M^{N_t}}\sum_k{\mathbb E}_{\mathbf n}\left( \log_2\left(\sum_{m}\exp (-\frac{1}{\sigma^2}||{\mathbf n}-{\mathbf H}{\mathbf G}({\mathbf x}_k-{\mathbf x}_m)||^2)\right) \right)\\
&=\frac{N_r}{\log(2)} +\frac{1}{M^{N_t}}\sum_{k}\int_{-\infty}^{+\infty} {\cal N}_c({\mathbf n}|{\mathbf 0},\sigma^2 {\mathbf I})
\log_2\left(\sum_{m}\exp (-\frac{1}{\sigma^2}||{\mathbf n}-{\mathbf H}{\mathbf G}({\mathbf x}_k-{\mathbf x}_m)||^2)\right) d{\mathbf n},
\label{eq_first}
\end{split}
\end{equation}
where ${\cal N}_c({\mathbf n}|{\mathbf 0},\sigma^2 {\mathbf I})$ represents the probability density function (pdf) of the circularly symmetric complex random vector due to AWGN. Let us define
\begin{equation}
\begin{split}
f_k \doteq \int_{-\infty}^{+\infty} {\cal N}_c({\mathbf n}|{\mathbf 0},\sigma^2 {\mathbf I})
\log_2\left(\sum_{{\mathbf x}_m}\exp (-\frac{1}{\sigma^2}||{\mathbf n}-{\mathbf H}{\mathbf G}({\mathbf x}_k-{\mathbf x}_m)||^2)\right) d{\mathbf n} \label{eq_special}.
\end{split}
\end{equation}
There is an strong connection between $f_k$ and the parameter
$H_k({\mathbf y})\doteq {\mathbb E}_{{\mathbf y}|{\mathbf x}_k}\left\{-\log_2(p({\mathbf y}))|{\mathbf x}={\mathbf x}_k \right\}$ called the Input-Dependent Output Entropy (IDOE) herein. Note that IDOE represents the entropy at the receiver output when the input is ${\mathbf x}_k$. This parameter is different that the conditional entropy. By using standard entropic identities, we can easily see that
\begin{equation}
\begin{split}
f_k = -H_k({\mathbf y})  + 2N_r\log_2(\sigma) + N_r\log_2(\pi) +N_t\log_2(M). \label{IDOE}
\end{split}
\end{equation}
Thus, since our Gauss-Hermite approximation focuses on finding approximations to each of the $f_k$ terms, one per input symbol, it equivalently offers estimates of the IDOE terms $H_k({\mathbf y})$. This gives a physical meaning to the estimated terms ${\hat f}_k$ presented below.

Since ${\mathbf n}$ has independent components over the different receiving antennas, and over the real and imaginary dimensions, the integral above can be partitioned into $2N_r$ real integrals in tandem, in the following manner: Define by $n_{rv}, n_{iv}$, with $v=1,\cdots,N_r$, the $v$th receiving antenna real and imaginary noise component, respectively. Also define by $({\mathbf H}{\mathbf G}({\mathbf x}_k-{\mathbf x}_m))_{rv}$ and $({\mathbf H}{\mathbf G}({\mathbf x}_k-{\mathbf x}_m))_{iv}$, the $v$th receiving antenna real and imaginary component of $({\mathbf H}{\mathbf G}({\mathbf x}_k-{\mathbf x}_m))$, respectively. We then have
\begin{equation}
{\cal N}_c({\mathbf n}|{\mathbf 0},\sigma^2 {\mathbf I}) = \frac{1}{\pi^{N_r}\sigma^{2N_r}}\exp(-\frac{\sum_l n_{rv}^2 + n_{iv}^2}{\sigma^2}),
\end{equation}
\begin{equation}
d{\mathbf n} = \prod_{v=1}^{N_r}dn_{rv}dn_{iv},
\end{equation}
and
\begin{equation}
\begin{split}
&\sum_{m}\exp (-\frac{1}{\sigma^2}||{\mathbf n}-{\mathbf H}{\mathbf G}({\mathbf x}_k-{\mathbf x}_m)||^2) \\
&=\sum_{m}\exp (-\frac{1}{\sigma^2}(\sum_v({ n}_{rv}-({\mathbf H}{\mathbf G}({\mathbf x}_k-{\mathbf x}_m))_{rv})^2\\
&+\sum_v({ n}_{iv}-({\mathbf H}{\mathbf G}({\mathbf x}_k-{\mathbf x}_m))_{iv})^2)).
\end{split}
\end{equation}
Based on the above equations, the $N_r$-size complex integral of (\ref{eq_special}) can be written as a $2N_r$-size real integral to which the GH quadrature approximation can be applied easily.

By applying the GH quadrature theory to the integral of a Gaussian function multiplied with an arbitrary real function $f(x)$, i.e.,
\begin{equation}
F \doteq \int_{-\infty}^{+\infty}\exp(-x^2)f(x)dx,
\end{equation}
one gets the following  approximation upon employing with $L$ GH weights and GH nodes as
\begin{equation}
F \approx \sum_{l=1}^L w(l) f(v_l) = {\mathbf w}^T {\mathbf f},
\end{equation}
with ${\mathbf w}=[w(1) \cdots w{(L)}]^t$, $\{v_l\}_{l=1}^L$, and ${\mathbf f} = [f(v_1) \cdots f(v_L)]^t,$ being the vector of the GH weights, the GH nodes, and the function GH node values, respectively. For the GH weights and nodes we have  \cite{G_H}
\begin{equation}
w(l) = \frac{2^{L-1}L! {\sqrt2\pi}}{L^2 (H_{L-1}(v_l))^2}
\end{equation}
where $H_{L}(x) = (-1)^{L} \exp(x^2) \frac{d^{L}}{dx^{L}}(\exp(-x^2))$ is the $(L)$-th order Hermitian polynomial, and the value of the node
$v_l$ equals the $l$th root of $H_L(x)$ for $l=1,2,\cdots,L$.
Let us first introduce some notations that make the overall understanding easier. Let ${\mathbf n}_e$ denote the equivalent to ${\mathbf n}$, complex vector of length $N_r$ derived from ${\mathbf n}$ as follows
\begin{equation}
{\mathbf n}_e = [n_{r1} +jn_{i1}\cdots n_{rN_r}+jn_{iN_r}]^T,
\end{equation}
with $n_{rv}+jn_{iv}$ being the values of the of the $v$th ($1\leq v\leq N_r$) element of ${\mathbf n}$, respectively. Let us also define the length $N_r$ complex vector defined as follows
\begin{equation}
{\mathbf v} = [v_{kr1}+jv_{ki1},\cdots , v_{krN_r}+jv_{kiN_r}]^T,
\end{equation}
with $k_{rv},~k_{iv}$ being in the set $\{1,2,\cdots,L\}$, i.e., ${\mathbf v}$ is a function of the complex vector ${\mathbf k}_c=[k_{r1}+jK_{i1}~ k_{r2}+jk_{i2}\cdots~k_{rN_r}+jk_{iN_r}]^T$.

Finally, the following lemma is proven in \cite{TK_EA_QAM} concerning the Gauss-Hermite approximation for $I({\mathbf x};{\mathbf y})$ in (\ref{eq_1}).\newline
\noindent{\bf Lemma.}
{\it For the MIMO channel model presented in (\ref{eq_1}), the Gauss-Hermite approximation for the input$I({\mathbf x};{\mathbf y})$ with $L$ nodes per receiving antenna is given as
\begin{equation}
\begin{split}
&I({\mathbf x};{\mathbf y})  \approx N_t\log_2(M) -\frac{N_r}{\log(
2)} -\frac{1}{M^{N_t}}\sum_{k=1}^{M^{N_t}}{\hat f}_k,\\
\label{eq_PAPER}
\end{split}
\end{equation}
where
\begin{equation}
\begin{split}
{\hat f}_k =&  \left(\frac{1}{\pi}\right)^{N_r}\sum_{k_{r1}=1}^{L} \sum_{k_{i1}=1}^{L}\cdots \sum_{k_{rN_r}=1}^{L} \sum_{k_{iN_r}=1}^{L}  w(k_{r1})w(k_{i1})\cdots w(k_{rN_r})\\
&\times w(k_{iN_r})z(\sigma {\mathbf v({\mathbf k}_c)}), \label{eq_f}
\end{split}
\end{equation}
with
\begin{equation}
z(\sigma {\mathbf v}({\mathbf k}_c))
\end{equation}
being the value of the function
\begin{equation}
\log_2\left(\sum_{m}\exp (-\frac{1}{\sigma^2}||{\mathbf n}-{\mathbf H}{\mathbf G}({\mathbf x}_k-{\mathbf x}_m)||^2)\right) \label{eq_basic}
\end{equation}
evaluated at ${\mathbf n}_e =\sigma{\mathbf v}({\mathbf k}_c)$.}

\section{Derivation of the MAI Interference Powers for JSDM-FA}
Here we derive the MAI for different precoding scenarios, including ZFP, ZF-PGP, and VAAC-PGP. In the following analysis we use the following generic notation for brevity.  ${\mathbf { H}}_{v,k,l,l'}$ denotes the effective virtual domain downlink channel for group's $k$, $l$ sub-group, containing the VCMBs employed by $G_{k,l'}$, and ${\mathbf P}_{v,k,l}$ is the precoder employed for the $G_{k,l}$ sub-group, based on a specific, but otherwise arbitrary technique employed for downlink precoding to sub-group $G_{k,l}$. Assume that we need to determine sub-group $G_{k,l'}$ interfering power to sub-group $G_{k,l}$, under JSDM-FA. Let the receiving equation for sub-group $G_{k,l}$ with interference from all sub-groups $l'\neq l$ be
\begin{equation}
{\mathbf y}_{k,l} = {{\mathbf H}}_{v,k,l,l}^H  {\mathbf P}_{v,k,l} {\mathbf x}_{k,l} + {{\mathbf n}}_{k,l,AWGN} + \sum_{l' ~\text{in}~G_{k,l},~l' \neq l}{{\mathbf n}}_{MAI,k,l,l'},
\end{equation}
where ${\mathbf y}_{k,l},~{\mathbf x}_{k,l},~{{\mathbf n}}_{k,l,AWGN},~\text{and}~{{\mathbf n}}_{MAI,k,l,l'}$ represent the ${\mathbf G}_{k,l}$ JSDM-FA sub-group received vector, data vector, AWGN noise vector, and MAI vector from ${\mathbf G}_{kl'}$ to ${\mathbf G}_{kl}$, respectively. We thus see that the covariance matrix of the interference from ${\mathbf G}_{kl'}$ to ${\mathbf G}_{kl}$, ${\mathbf K}_{k,l,l'}$ becomes
\begin{equation}
{\mathbf K}_{k,l,l'}={{\mathbf H}}_{v,k,l,l'}{\mathbf P}_{v,k,l'}{\mathbf P}_{v,k,l'}^{H}{{\mathbf H}}_{v,k,l,l'}^H,
\end{equation}
where the input symbols are uncorrelated and of unit power. Applying this result using the generic ${\mathbf P}_{v,k,l'}$ and focusing on the diagonal elements of ${\mathbf K}_{k,l,l'}$ only, we get the following expressions for the MAI power at the input of $G_{k,l}$ $i$th UE receiver from $G_{k,l'}$
\begin{equation}
{\mathbf p}_{MAI,k,l,l'}(i)={\mathbf K}_{k,l,l'}[i,i]
, \label{app}
\end{equation}
where $i=1,\cdots,N_{S_k}$. The total MAI to $G_{k,l}$ $i$th UE receiver is then
\begin{equation}
{\mathbf p}_{MAI,k,l}(i)=\sum_{l' ~\text{in}~G_{k},~l' \neq l}{\mathbf p}_{MAI,k,l,l'}(i).
\end{equation}
Then, for the two main precoder types considered in this paper, we use
\begin{equation}
{\mathbf P}_{v,k,l'}^{(ZFP)}= w_{ZF,k,l'}^2{\mathbf H}_{v,k,l'}^H\left({\mathbf H}_{v,k,l'} \cdot {\mathbf H}_{v,k,l'}^H  \right)^{-1},
\end{equation}
where $w_{ZF,k,l'}$ is the ZFP $\mathrm{SNR}$ normalizing weight of the ${\mathbf G}_{k,l}$ group \cite{ZF-PGP}, while for the ZF-PGP one, we get
\begin{equation}
{\mathbf P}_{v,k,l'}^{(ZF-PGP)}= {\mathbf S}_{EFF}{\mathbf {\tilde S}}_{v,k,l'}^{(ZF-PGP)}{\mathbf V}_{v,k,l'}^{(ZF-PGP)},
\end{equation}
where ${\mathbf S}_{EFF}$ is the effective channel singular value matrix in ZF-PGP \cite{ZF-PGP}, and ${\mathbf {\tilde S}}_{v,k,l'}^{(ZF-PGP)}=\sqrt{2}{\mathbf I}_{N_g}$ is the PGP singular values of the ZF-PGP precoder, due to VAAC, and ${\mathbf V}_{v,k,l'}^{(ZF-PGP)}$ is the right singular matrix of the ZF-PGP precoder.

\section{VAAC-PGP Derivation Details}
Here the concept of adding virtual antennas, i.e., additional data streams to the same antennas employed by a MIMO system jointly with PGP-WG is explained in detail. Without a loss of generality, we consider a MIMO system with equal number of transmitting and receiving antennas, i.e., $N_t=N_r=N_g$, where $N_t$ and $N_r$ represent the number of transmitting and receiving antennas, respectively. The channel model under consideration then becomes
\begin{equation}
{\mathbf y}={\mathbf H}{\mathbf x}+{\mathbf n},
\end{equation}
where ${\mathbf y},~{\mathbf H},~{\mathbf x},~\text{and}~{\mathbf n}$ represent the received data, the MIMO channel, the transmitted data, and the AWGN noise, respectively, and where  matrices are of size $N\times N$ and vectors are of size $N\times 1$. The equivalent singular value decomposition based model for the MIMO channel is
\begin{equation}
{\mathbf y}={\mathbf U}_H{\boldsymbol \Sigma}_H{\mathbf V}_H^H{\mathbf x}+{\mathbf n}, \label{eq_MIMO}
\end{equation}
with ${\mathbf U}_H,~{\boldsymbol \Sigma}_H,~{\mathbf V}_H$ representing the size $N\times N$ matrices of left singular vectors, singular values, and right singular vectors, respectively.
Consider adding $N$ virtual antennas of zero singular values, i.e., useless, noise-only channels. This can be added to the previous model as follows

\begin{eqnarray}
\begin{aligned}
\left[\begin{array}{c  } {\mathbf y}\\
{\mathbf y}_a\end{array} \right]&=&
\left[\begin{array}{c c } {\mathbf U}_H & {\mathbf 0}\\
{\mathbf 0} & {\mathbf I}_N\end{array} \right]
\left[\begin{array}{c c } {\boldsymbol  \Sigma}_H & {\mathbf 0}\\
{\mathbf 0} & {\mathbf 0}\end{array} \right]
\left[\begin{array}{c c } {\mathbf V}_H^H & {\mathbf 0}\\
{\mathbf 0} & {\mathbf I}_N\end{array} \right]
\left[\begin{array}{c  } {\mathbf x}\\
{\mathbf x}_a\end{array} \right] +
\left[\begin{array}{c  } {\mathbf n}\\
{\mathbf n}_a\end{array} \right], \label {eq_VAAC}
\end{aligned}
\end{eqnarray}
where the subscript $a$ is used to indicate the $N$ added, fictitious antennas. In the above equation, the vector ${\mathbf x}_a$ represents the $N$ added QAM inputs to the MIMO system. Note that in (\ref{eq_VAAC}) the inputs represented by ${\mathbf x}_a$ cannot be transmitted, due to their corresponding zero singular values (noise-only channel) which result in zero input-output mutual information. However, one can still apply the virtual model of (\ref{eq_VAAC}) with PGP-WG. The PGP-WG algorithm will optimize and assign an amplitude diagonal matrix as per PGP-WG ${\boldsymbol \Sigma}_{P_i}=\mathrm{diag}[\sqrt{ 2}, ~0]$, $i=1,2,\cdots,K_g$ for each sub-group of the PGP-WG, i.e., no power sent to the noise-only antenna. This results in ${\mathbf S}_P=\sqrt{2}{\mathbf I}_{N_g}$ in VAAC-PGP. On the other hand, PGP-WG will also determine the optimal unitary precoder matrix to each sub-group in PGP-WG, thus it will be multiplexing optimally two QAM symbols to each actual transmitting antenna of the original MIMO system in (\ref{eq_MIMO}). Now, VAAC-PGP proceeds as follows: for group $g$, $g=1,2,3,4$, the precoder employs the singular values of the downlink channel ${\mathbf H}_{g,v}$ as the matrix ${\mathbf \Sigma}_H$ of equation (\ref{eq_VAAC}). Then, it applies PGP-WG as explained above. VAAC-PGP is basically similar to ZF-PGP, but without the ZF part, i.e., no left diagonalization of the channel matrix takes place in the BS precoding process, thus the need for the UEs to do that arises.

\section{Proof of Theorem 1 and the Corollaries of OPGPA}
First, we prove Theorem 1. By multiplying (\ref{eq_OPGPA}) by $\frac{1}{\sqrt {2k_m} s_{g,v,eff,m}^{(\mathrm{SNR}_0)}}$, we get (\ref{eq_OPGPA_1}). Since the original circular complex Gaussian noise vector in (\ref{eq_OPGPA}) has a variance 1 per component (due to normalization) with correlation 0 (independent components), the new noise vector has a covariance matrix equal to
\begin{equation}
\frac{1}{2{k_m}\left(s_{g,v,eff,m}^{(\mathrm{SNR}_0)} \right)^{2}}{\mathbf I}_2.
\end{equation}
Thus, the effective $\mathrm{SNR}$ of the PGP group is $\mathrm{SNR}_{eff,m}=2k_m \left(s_{g,v,eff,m}^{(\mathrm{SNR}_0)} \right)^2$. To prove the rest of Theorem 1 claims, we first notice that the IOMIM precoder operating on (\ref{eq_OPGPA}) depends only on the effective $\mathrm {SNR}$. Thus, the IOMIM precoder for the $m$ PGP group is the same, under the requirement that $\mathrm{SNR}_{eff,m}=\mathrm{constant}$ for all $m=1,2,\cdots,N_g$.

For Corollary 1, we notice that by setting $k_m=1$ in (\ref{eq_OPGPA}), we get $\mathrm{SNR}_{eff,m}=\frac{1}{2\left(s_{g,v,eff,m}^{(\mathrm{SNR}_0)} \right)^{2}}$, thus the claim becomes obvious.

For Corollary 2, we use the model in (\ref{eq_OPGPA_1}) and observe that for a set $\mathrm{SNR}_{eff,m}$ the precoder is constant, in other words, the IOMIM precoder is only a function of $\mathrm{SNR}_{eff,m}$. Since an increase of $\mathrm{SNR}_{eff,m}$ results in higher throughput by the IOMIM precoder, we see that there is a unique required $\mathrm{SNR}_{req}(I_S)$ for each $I_S$. This means that by setting $\mathrm{SNR}_{eff,m}=\mathrm{SNR}_{req}(I_S)$, the desired $I_S$ is attained by all PGP groups.

Finally, for Corollary 3, we need to substitute into $\mathrm{SNR}_{eff,m}=\mathrm{SNR}_{req}(I_S)$ the expression $\mathrm{SNR}_{eff,m}=2k_m \left(s_{g,v,eff,m}^{(\mathrm{SNR}_0)} \right)^2$ from Theorem 1. Then, we get directly that the $m$th ($1\leq m \leq N_g$) gain is determined as follows $\sqrt{k_m} = {\sqrt \frac{\mathrm{SNR}_{req}(I_S)}{2}}\frac{1}{s_{g,v,eff,m}^{\mathrm{(\mathrm{SNR}_0)}}}$.


\end{document}